\documentclass{article}
\usepackage{amsmath} 
\usepackage{epsfig} 
\usepackage{cite} 

\input{epsf} 
\def\istring#1#2{\mbox{$({#1}_1,\dots,{#1}_{#2})$}} 
\def\la{\langle}
\def\ra{\rangle}
\def\ve{\vert}

\def\b0{b_0}

\def\beq{\begin{equation}} 
\def\eeq{\end{equation}} 
\def\beeq{\begin{eqnarray}} 
\def\eeeq{\end{eqnarray}}

\def\nn{\nonumber}

\def\gev2{{\rm GeV}^2}
\def\nn{\nonumber}

\def\lapproxeq{{\ \lower 0.6ex \hbox{$\buildrel<\over\sim$}\ }}
\def\gapproxeq{{\ \lower 0.6ex \hbox{$\buildrel>\over\sim$}\ }}

\def \be  {\begin{equation}}
\def \ee  {\end{equation}}
\def \ba  {\begin{eqnarray}}
\def \ea  {\end{eqnarray}}
\def \baa {\begin{eqnarray*}}
\def \eaa {\end{eqnarray*}}
\def \bb  {\begin{thebibliography}}
\def \eb  {\end{thebibliography}}

\begin{document} 

\begin{titlepage}
\renewcommand{\thefootnote}{\fnsymbol{footnote}}
\begin{flushright}
     \end{flushright}
\par \vspace{10mm}

\begin{center}
{\Large \bf
The last of the seven-parton tree amplitudes}
\end{center}
\par \vspace{2mm}
\begin{center}
{\bf Daniel de Florian\footnote{email address: deflo@df.uba.ar} and 
Jos\'e Zurita\footnote{email address: jzurita@df.uba.ar}}\\

\vspace{5mm}

Departamento de F\'\i sica, FCEYN, Universidad de Buenos Aires,\\
(1428) Pabell\'on 1 Ciudad Universitaria, Capital Federal, Argentina\\
\vspace{2mm}

\vspace{5mm}

\end{center}

\par \vspace{2mm}
\begin{center}

 {\large \bf Abstract}\\
\vspace{1cm}

We compute the four-quark plus three-gluon and six-quark plus one-gluon tree level amplitudes 
using on-shell recursion relations. They are needed for the calculation of  the 5-jet 
cross-section at the Born level, and constitute an essential ingredient for next-to-leading 
order 4-jet and next-to-next-to-leading order 3-jet production at hadronic colliders. Very 
compact expressions for all possible helicity configurations are provided, allowing direct 
implementation in computer codes. With the results presented in this paper, the full set of 
seven-parton tree amplitudes becomes available.

 \end{center}
\begin{quote}
\pretolerance 10000
                    
\end{quote}

\vspace*{\fill}
\begin{flushleft}
August 2006

\end{flushleft}
\end{titlepage}

\setcounter{footnote}{1}
\renewcommand{\thefootnote}{\fnsymbol{footnote}}

\section{Introduction}
\label{sec:intro}

With the forthcoming of the Large Hadronic Collider (LHC), the possibility to explore a wide range 
of new phenomena, like Higgs production, dark matter candidates, SUSY and new quark flavors will 
open. In order to study new physics, accurate calculations of cross-sections involving many 
partons are needed, for both signal and background. Such computation involves 
typically of the order of a thousand Feynman diagrams, so many that traditional methods become 
obsolete. During the 80's, some new ideas were introduced 
\cite{Mangano:1990by,Dixon:1996wi,Xu:1986xb,
Kleiss:1985yh,Gunion:1985vc,Berends:1987cv,Mangano:1987xk} to allow a great simplification of the 
calculus, making possible the implementation of automatic computer algorithms (like MadEvent 
\cite{Maltoni:2002qb}) to perform this task. However, that still implies a huge computational 
effort, and the calculation of  cross-sections for processes involving many partons remains as 
an unsolved problem.

In the last few years, the situation has drastically changed. Witten first noted the duality 
between tree level amplitudes and twistor string theory \cite{Witten:2003nn}. This lead  to 
the introduction of the off-shell recursion relations by Cachazo, Svr\v{c}ek and Witten 
\cite{Cachazo:2004kj} (CSW), allowing to compute amplitudes for several partons by iterating the 
simplest  MHV (Maximally Helicity Violating) ones \cite{Parke:1986gb}. Britto, 
Cachazo and Feng presented, as an ansatz, the on-shell recursion relations \cite{Britto:2004ap}, 
which were later proved to be correct (BCFW) \cite{Britto:2005fq}. The method was found very 
useful to provide compact expressions 
for gluon amplitudes. Soon it was extended to include massless fermions 
\cite{Georgiou:2004wu,Georgiou:2004by}, Higgs boson \cite{Dixon:2004za}, massive gauge bosons 
\cite{Bern:2004ba}, photons \cite{Ozeren:2005mp} and massive fermions \cite{Badger:2005zh}.

At tree level, six-parton processes were studied long time ago 
\cite{Gunion:1986zg,Gunion:1986zh}, and many of the amplitudes have been recomputed using the 
recursion relations recently \cite{Luo:2005rx,Luo:2005my}. Seven-gluon amplitudes were first 
calculated in \cite{Bern:2004ky}, while certain eight-gluon results can be found in 
\cite{Britto:2004ap,Roiban:2004ix}. The two-fermion plus five-gluon amplitudes were presented in 
our previous paper \cite{deFlorian:2006ek}. At the loop level, the recursion relations become 
more complicated. However, calculations have reached the level of one-loop six-gluon amplitudes, 
and also some results for certain helicity configurations with 
an arbitrary number of partons were presented 
\cite{Bern:2005hs,Bern:2005ji,Bern:2005cq,Bern:2005hh,Bern:2006dp,Berger:2006ci,
Bjerrum-Bohr:2006rz,Berger:2006vq,Xiao:2006vr,Su:2006vs,Xiao:2006vt,Badger:2006us}.

Recent work  \cite{Duhr:2006iq,Dinsdale:2006sq} also focused on the numerical implementation of 
the on-shell recursion relations. The result turns out to be that for a large number of partons, 
the ``old'' Berends-Giele \cite{Berends:1987cv} recursion relation performs better 
than the twistor inspired methods. Nevertheless, counting with analytical results for non 
trivial amplitudes would certainly speed up numerical calculations, reducing 
considerably the CPU-time required by any
automatic algorithm. Therefore, obtaining very simple and compact results for the 
amplitudes is strongly desirable.
   
In this paper we present all four-quark plus three-gluon 
and six-quark plus one-gluon tree level helicity amplitudes, searching always for their most 
compact expressions. Our results have been checked against the factorization properties of the 
amplitudes, which provides a very strong consistency test. 
Combined with previous calculations \cite{Britto:2004ap,deFlorian:2006ek}, the full set of 
seven-parton tree level amplitudes is now available.

This article is organized as follows: in Section 2 we review the color decomposition of 
tree level processes, and the main ingredients of the BCFW formulation, recalling its 
limitations when dealing with fermions. In Sections 3 and 4 we introduce the 
main results for the complete set of seven-parton amplitudes with two and three quark-
antiquark pairs, respectively, while in Section 5 we 
present our conclusions. The necessary six-parton formulae for our calculation are 
collected in the appendix.

\section{Color decomposition, helicity and BCFW}
\label{sec:review}
 
As is well known in the QCD literature, the color factor for the $m$ $q \bar{q}$ pair plus 
$n$ gluon amplitude is \cite{Mangano:1990by}

\beq    \label{qgcolor}       
\Lambda (\{n_i\}, \{\alpha\}) \,  = 
\frac{(-1)^p}{N^{p}} (T^{a_1} \dots T^{a_{n_1}})_{i_1 \alpha_1}
(T^{a_{n_1+1}} \dots T^{a_{n_2}})_{i_2\alpha_2}    
\dots(T^{a_{n_{m-1}+1}}\dots T^{a_{n}})_{i_m\alpha_m}
\eeq 
where $i_1,\dots,i_m$ are the color indices of the quarks and 
$\alpha_1,\dots,\alpha_m$ are the ones for the antiquarks. 
The indices $n_1,\dots,n_{m-1}$ (with $1\le n_i\le n$) correspond to an arbitrary 
partition
of an arbitrary permutation of the $n$ gluon indices.  
Each external quark is connected by a fermionic line to an 
external antiquark. When we want to
indicate that a quark with color index $i_k$ and an antiquark with color
index $\alpha_k$ are in fact connected by a fermionic line, we identify the
index $\alpha_k$ with the index $\bar{i}_k$. Then the string
$\{\alpha\}=\istring{\alpha}{m}$ is a generic permutation of the string
${\bar{i}}=\istring{\bar{i}}{m}$.  The power $p$ is determined by 
the number of times $\alpha_k = \bar{i}_k$.   If  ${\alpha}\equiv
{\bar{i}}$, then $p=m-1$.  
As usual,  a product of zero $T$ matrices has to be understood as a Kronecker delta.

The full amplitude typically reads
\beq
M_{n+m}^{(0)}=\sum_{\{n_k\}, \{\alpha_k\}} \Lambda (\{n\}, \{\alpha\})
A_{n+m}^{(0)} ( \{i\}, \{n\}, \{\alpha\}) \, ,
\eeq
where the $A_{n+m}^{(0)}$ are known as the \emph{partial amplitudes}. They depend only 
on the kinematic of the process, namely, the momenta and helicities of the particles.

In the framework of the helicity formalism 
\cite{Xu:1986xb,Kleiss:1985yh,Gunion:1985vc}, with the spinors denoted as 
\begin{equation}
\label{spinors}
\vert i^{\pm}\rangle=\vert k_i^{\pm}\rangle=\psi_{\pm} (k_i) \qquad \qquad 
\langle i^{\pm}\vert=\langle k_i^ {\pm}\vert=\overline{\psi_{\pm} (k_i)} \, ,
\end{equation}
the partial amplitudes can be written in terms of the spinors inner-products
\begin{equation}
\label{inner}
\begin{split}
&\langle ij\rangle=\langle i^-\vert j^+\rangle=\overline{\psi_{-}}(k_i) \psi_{+} (k_j) \\
&[ij]=\langle i^+ \vert j^-\rangle=\overline{\psi_{+}} (k_i) \psi_{-} (k_j) \, ,
\end{split}
\end{equation}
and a few simple combinations of them, like
\begin{eqnarray}
\langle i \vert p_a\vert j] \equiv \langle i a \rangle [a j] \nn \\
\langle i \vert p_a p_b\vert j\rangle 
\equiv \langle i a \rangle [a b]\langle b j \rangle  \, .
\end{eqnarray}
In our convention all particles are considered to be outgoing and, following 
the QCD literature \cite{Mangano:1990by,Dixon:1996wi}, we fix the sign of the 
inner products such that  $\langle ij \rangle [ji] = s_{ij}$
\footnote{When comparing with results obtained using the 
string-like conventions just notice that $[ij]$ carries the opposite sign}.

The BCFW recurrence relation is based on the analytical properties of the amplitude 
when the spinors of two external legs  (denoted by $j$ and $l$) are shifted as 
\begin{equation}\label{bcfseg}
\begin{split}
&\vert \hat{j} \rangle = \vert j \rangle  \\
&\vert \hat{j}]= \vert j] - z \vert l] \\
&\vert \hat{l} \rangle = \vert l \rangle + z \vert j \rangle \\
&\vert \hat{l}]= \vert l ] \,.
\end{split}
\end{equation}
After this shift, the BCFW formula simply reads
\begin{equation}\label{BCF}
\begin{split}
A_n^{(0)} (1^{\lambda_1},\dotsc,n^{\lambda_n})= \sum_{r,s}\sum_{\lambda =\pm} 
A_{s-r+2}^{(0)} (r^{\lambda_r},\dotsc,
\hat{j}^{\lambda_j},\dotsc,s,-\hat{K}_{r s}^{\lambda}) \\ \times \,
\frac{1}{K^2_{r s}}\, A_{n-s+r}^{(0)} (\hat{K}_{r s}^{-\lambda},(s+1)^{\lambda_{(s+1)}},
\dotsc,\hat{l}^{\lambda_l},\dotsc,(r-1)^{\lambda_{(r-1)}}) \, , 
\end{split}
\end{equation}
where $K_{rs}= k_r +\dotsc+k_j \dotsc+k_s $ and the (complex) shift variable $z$ 
takes the value
\begin{equation}
\label{zrs}
z_{rs}=-\frac{(K_{rs})^2}{\langle j \vert K_{rs}\vert l]} \, .
\end{equation}

The sum on Eq.(\ref{BCF}) is over all the possible configurations where the $j$ 
particle belongs to one of the amplitudes and the $l$ particle to the other one. It 
includes  also the 3-particle sub-amplitudes, which are usually vanishing for 
on-shell particles 
but become non zero because of the BCFW shift in Eq.(\ref{bcfseg}). 

When tagging the particles, one has to fulfill certain restrictions, 
i.e.: some configurations of $j$ and $l$ are forbidden.
The helicities of the reference particles can not be chosen as 
$(\lambda_j,\lambda_l)=(+,-)$. 

Quarks and antiquarks of the same flavor can not be selected if they are 
adjacent. Furthermore, one can not chose two adjacent fermions of different 
flavor and the same helicity.
For adjacent quarks and gluons the helicities should better be opposite. 
\cite{Luo:2005rx}.

Under a general tag, some sub-amplitudes may vanish. A clever choice of $j$ and $l$ leads to 
most compact expressions for the same amplitudes. In this paper, we have chosen the allowed tag 
that gives the minor number of terms. 
Nevertheless, in some calculations we have tried some forbidden tags, 
in the search for results with 
fewer terms and simpler structure. In that case we explicitly checked that the final 
expressions 
agreed with the longer forms obtained using allowed legs in the BCFW recursion relation.

The main ingredient of the recursion relations, that permits  to construct the tree 
level amplitude for $n$-partons just by iterations, are the $ggg$ and $q\bar{q}g$ 
MHV amplitudes \cite{Parke:1986gb},  which with our phase conventions read
\begin{eqnarray}
\label{3amp}
A _3^{(0)} (1^{+}_g,2^{-}_g,3^{-}_g)= \frac{\la 2 3\ra^3}
{\langle 1 2\rangle \langle 3 1\rangle} \, , \nn \\
A _3^{(0)} (1^{+}_q,2^{-}_{\bar{q}},3^{-}_g)= \frac{\la 2 3\ra^2}{\la 2 1\ra }  \, .
\end{eqnarray}
The corresponding $\overline{\rm MHV}$ amplitudes can be obtained from those above by 
using some simple properties. 
The ``parity'' operation inverts the helicities of all the particles, and at the level of the 
amplitudes implies to perform the exchange 
$\langle i j\rangle \leftrightarrow [ji]$ 
and add an extra factor of (-1) for each fermionic pair involved. 
Charge conjugation turns quarks into antiquarks 
and vice-versa. The partial amplitudes are C-invariant: they do not change under 
the action of this operator. Also one can profit from the reflection and cyclic symmetries 
of the amplitudes. The reflection property states that the following equality holds
\beq
A(1,2,\dots,n-1,n)=(-1)^n A(n,n-1,\dots,2,1) \, ,
\eeq
while cyclic symmetry implies that the amplitude is invariant under cyclic 
permutations of $1,2,\dots,n$. 
The factorization properties of the amplitudes in the collinear and soft channels 
are an useful tool to check for the consistency of the results.
Further details on the method of calculation  can be found in \cite{deFlorian:2006ek}.

\section{Four-quark plus three-gluon amplitudes}

Seeking for the most general presentation, we concentrate on the case of different quark 
flavors, that we denote as $q$ and $p$. To obtain the equal flavor two-quark pair 
amplitudes, it is enough to substract from a given partial amplitude the same term with the 
antiquarks exchanged (and not the quarks) \cite{Gunion:1986zg, DelDuca:1999pa}.

From Eq.(\ref{qgcolor}), we can read off the two different color structures contributing to 
the two-quark pair plus gluon amplitudes. For $p=0$, it follows 
$\alpha_1=\bar{i}_2, \alpha_2=\bar{i}_1 $ and one obtains the leading 
contribution
\beq
(T^{a(\sigma_1)} \dots T^{a(\sigma_j)})_{i_1 \bar{i}_2} 
(T^{a(\sigma_{j+1})} \dots T^{a(\sigma_l)})_{i_2 \bar{i}_1} \, ,
\eeq
while if $p=1$ (in this case ${\alpha}\equiv {\bar{i}}$) there is a
sub-leading term
\beq
\frac{-1}{N} (T^{a(\sigma_1)} \dots T^{a(\sigma_j)})_{i_1 \bar{i}_1} 
(T^{a(\sigma_{j+1})} \dots T^{a(\sigma_l)})_{i_2 \bar{i}_2} \, .
\eeq
For leading color, the partial 
amplitude corresponds to the following string of quarks and gluons
\beq\label{lc}
[q (g_{\sigma_1} \dots g_{\sigma_j}) \bar{p}] 
[p (g_{\sigma_{j+1}} \dots g_{\sigma_l}) \bar{q}] \, ,
\eeq
while for the sub-leading color the ordering is
\beq\label{slc}
 [q (g_{\sigma_1} \dots g_{\sigma_j}) \bar{q}] 
[p (g_{\sigma_{j+1}} \dots g_{\sigma_l}) \bar{p}] \, .
\eeq 
In order to maintain a similar notation (the first particle being always $q$) to the one used 
in \cite{deFlorian:2006ek} for 
one-quark pair amplitudes, we will present the results with the following format
\beq
A (q, \bar{q}, g_{\sigma_1}, \dots, g_{\sigma_j}, p, \bar{p}, 
g_{\sigma_{j+1}}, \dots, g_{\sigma_l}) 
\eeq
and
\beq
A (q, \bar{p}, g_{\sigma_1}, \dots, g_{\sigma_j},p, \bar{q}, g_{\sigma_{j+1}}, \dots, g_{\sigma_l})  \, ,
\eeq 
for the leading and sub-leading color structures, respectively. Those corresponding 
exactly to the ordering in Eq.(\ref{lc}) and Eq.(\ref{slc}) can be easily obtained from the ones
above by $C$ operation and relabelling of the partons under cyclic symmetry.

Several amplitudes are accounted for by the MHV formula for two-quark pairs 
and $n-4$ gluons, which is given by
\beq\label{4qmhv}
A_n^{(0)}(1_q,2_{\bar{q}},3_g,\dots,k_g, (k+1)_{p}, (k+2)_{\bar{p}},(k+3)_g,\dots,n_g)=
\frac{(-1) \la ij \ra^2 \la p \bar{q} \ra \la q \bar{p} \ra}
{\prod^n_{l=1} \langle l\medspace \medspace l+1\rangle} \, ,
\eeq
where $i$ and $j$ are the two fermions with negative helicity while all the 
other particles
have positive helicity \cite{Mangano:1990by}. We have reobtained Eq.(\ref{4qmhv}) 
using the BCFW recursion relation and Eq.(\ref{3amp}).
For the sub-leading case \emph{exactly} the same formula holds. 

When facing the NHMV case, we denote the leading and sub-leading color contributions 
as ``A'' and ``B'', respectively.
There are fourteen independent ``A'' amplitudes and twelve independent ``B'' amplitudes.
For simplicity, we will refer to them as 
$A_1, A_2, \dots, A_{14}$ and $B_1, B_2, \dots, B_{12}$.
The full list is
\beq\label{4q3gaamp}
\begin{split}
A_1= A_7^{(0)}(1_q^+,2_{\bar{q}}^-,3^+,4^+,5^-,6_p^+, 7_{\bar{p}}^-) \qquad \qquad
&A_8= A_7^{(0)}(1_q^+,2_{\bar{q}}^-,3^-,4^+,5_p^+, 6_{\bar{p}}^-,7^+) \\
A_2= A_7^{(0)}(1_q^+,2_{\bar{q}}^-,3^+,4^-,5^+,6_p^+, 7_{\bar{p}}^-) \qquad \qquad
&A_9= A_7^{(0)}(1_q^+,2_{\bar{q}}^-,3^+,4^+,5_p^-, 6_{\bar{p}}^+,7^-) \\
A_3= A_7^{(0)}(1_q^+,2_{\bar{q}}^-,3^-,4^+,5^+,6_p^+, 7_{\bar{p}}^-) \qquad \qquad
&A_{10}= A_7^{(0)}(1_q^+,2_{\bar{q}}^-,3^+,4^-,5_p^-, 6_{\bar{p}}^+,7^+) \\
A_4= A_7^{(0)}(1_q^+,2_{\bar{q}}^-,3^+,4^+,5^-,6_p^-, 7_{\bar{p}}^+) \qquad \qquad
&A_{11}= A_7^{(0)}(1_q^-,2_{\bar{q}}^+,3^+,4^+,5^-,6_p^+, 7_{\bar{p}}^-) \\
A_5= A_7^{(0)}(1_q^+,2_{\bar{q}}^-,3^+,4^-,5^+,6_p^-, 7_{\bar{p}}^+) \qquad \qquad
&A_{12}= A_7^{(0)}(1_q^-,2_{\bar{q}}^+,3^+,4^-,5^+,6_p^+, 7_{\bar{p}}^-) \\
A_6= A_7^{(0)}(1_q^+,2_{\bar{q}}^-,3^+,4^+,5_p^+, 6_{\bar{p}}^-,7^-) \qquad \qquad
&A_{13}= A_7^{(0)}(1_q^-,2_{\bar{q}}^+,3^+,4^+,5_p^+, 6_{\bar{p}}^-,7^-) \\
A_7= A_7^{(0)}(1_q^+,2_{\bar{q}}^-,3^+,4^-,5_p^+, 6_{\bar{p}}^-,7^+) \qquad \qquad
&A_{14}= A_7^{(0)}(1_q^-,2_{\bar{q}}^+,3^+,4^-,5_p^+, 6_{\bar{p}}^-,7^+) \\
\end{split}
\eeq
for type A processes and
\beq
\begin{split}
B_1= A_7^{(0)}(1_q^+,2_{\bar{p}}^-,3^+,4^+,5^-,6_p^+, 7_{\bar{q}}^-) \qquad \qquad
&B_7= A_7^{(0)}(1_q^+,2_{\bar{p}}^-,3^+,4^+,5_p^+, 6_{\bar{q}}^-,7^-) \\
B_2= A_7^{(0)}(1_q^+,2_{\bar{p}}^-,3^+,4^-,5^+,6_p^+, 7_{\bar{q}}^-) \qquad \qquad
&B_8= A_7^{(0)}(1_q^+,2_{\bar{p}}^-,3^+,4^-,5_p^+, 6_{\bar{q}}^-,7^+) \\
B_3= A_7^{(0)}(1_q^+,2_{\bar{p}}^-,3^-,4^+,5^+,6_p^+, 7_{\bar{q}}^-) \qquad \qquad
&B_9= A_7^{(0)}(1_q^+,2_{\bar{p}}^-,3^-,4^+,5_p^+, 6_{\bar{q}}^-,7^+) \\
B_4= A_7^{(0)}(1_q^+,2_{\bar{p}}^+,3^+,4^+,5^-,6_p^-, 7_{\bar{q}}^-) \qquad \qquad
&B_{10}= A_7^{(0)}(1_q^+,2_{\bar{p}}^+,3^+,4^+,5_p^-, 6_{\bar{q}}^-,7^-) \\
B_5= A_7^{(0)}(1_q^+,2_{\bar{p}}^+,3^+,4^-,5^+,6_p^-, 7_{\bar{q}}^-) \qquad \qquad
&B_{11}= A_7^{(0)}(1_q^+,2_{\bar{p}}^+,3^+,4^-,5_p^-, 6_{\bar{q}}^-,7^+) \\
B_6= A_7^{(0)}(1_q^+,2_{\bar{p}}^+,3^-,4^+,5^+,6_p^-, 7_{\bar{q}}^-) \qquad \qquad
&B_{12}= A_7^{(0)}(1_q^+,2_{\bar{p}}^+,3^-,4^+,5_p^-, 6_{\bar{q}}^-,7^+) \\
\end{split}
\eeq
for type B processes.

The explicits results for those amplitudes are presented in the next sub-sections.

\subsection{4q+3g leading amplitudes}
We present now the most compact expressions for the 14 independent leading amplitudes.

We have taken $j=5, l=6$ for the calculation of $A_1$, leading to
\beq
\begin{split}
A_1&= A_7^{(0)}(1_q^+,2_{\bar{q}}^-,3^+,4^+,5^-,6_p^+, 7_{\bar{p}}^-)= \\
&+\frac{\la 57 \ra^3 \la 2 \ve 3+1 \ve 4 ]^3 }{s_{567}s_{123} \la 12 \ra 
\la 23 \ra \la 67 \ra \la 7 \ve 6+5 \ve 4] \la 3 \ve (2+1) (7+6) \ve 5 \ra}\\
&+\frac{\la 57 \ra^3 \la 5 \ve 4+3 \ve 1 ]^3 }
{[12] \la 34 \ra \la 45 \ra \la 67 \ra \la 5 \ve 6+7 \ve 1] 
\la 3 \ve (2+1) (7+6) \ve 5 \ra  \la 5 \ve (4+3) (2+1) \ve 7 \ra} \\
&-\frac{[61]^2 \la 25 \ra^3 \la 5 \ve 1+7 \ve 6 ]} {s_{671} \la 23 \ra 
\la 34 \ra \la 45 \ra [67] \la 2 \ve 1+7 \ve 6] \la 5 \ve 6+7 \ve 1]} \\
&-\frac{\la 27 \ra^2 \la 5 \ve 4+3 \ve 6]^3 \la 5 \ve (4+3) (1+7) \ve 2 \ra}
{s_{712}s_{345} \la 12 \ra \la 34 \ra \la 45 \ra \la 3 \ve 4+5 \ve 6 ]
\la 2 \ve 1+7 \ve 6 ] \la 5 \ve (4+3) (2+1) \ve 7 \ra} \\
&+\frac{\la 27\ra^2 [46]^3 \la 2 \ve 6+5 \ve 4]}{s_{456} \la 12 \ra 
\la 23 \ra [45] [56] \la 3 \ve 4+5 \ve 6 ] \la 7 \ve 6+5 \ve 4] } \, .\\
\end{split}
\eeq

$A_2$ has been calculated by selecting $j=7, l=1$, obtaining
\beq
\begin{split}
A_2&= A_7^{(0)}(1_q^+,2_{\bar{q}}^-,3^+,4^-,5^+,6_p^+, 7_{\bar{p}}^-)= \\
&+\frac{\la 46 \ra \la 27 \ra^2 \la 2 \ve 1+7 \ve 3] \la 4 \ve 5+6 \ve 3]^3}
{s_{456}s_{712} \la 12 \ra \la 45 \ra \la 56 \ra \la 6 \ve 5+4 \ve 3 ]
\la 7 \ve 1+2 \ve 3 ] \la 2 \ve (1+7) (6+5) \ve 4 \ra } \\
&+\frac{\la 27 \ra^2 \la 24 \ra^4 [56]^2 \la 2 \ve 1+7 \ve 5]}
{\la 12 \ra \la 23 \ra \la 34 \ra \la 2 \ve 3+4 \ve 5] \la 2 \ve 1+7 \ve 6]
\la 7 \ve (6+5) (4+3) \ve 2 \ra \la 2 \ve (1+7) (6+5) \ve 4 \ra } \\
&-\frac{\la 26 \ra \la 27 \ra^2 [35]^4}{s_{345} \la 12 \ra [34] [45] 
\la 67 \ra \la 2 \ve 3+4 \ve 5] \la 6 \ve 5+4 \ve 3]}\\
&-\frac{\la 46 \ra \la 47 \ra^3 [13]^3}{s_{123} [12] \la 45 \ra \la 56 \ra 
\la 67 \ra \la 4 \ve 3+2 \ve 1] \la 7 \ve 1+2 \ve 3]}\\
&-\frac{\la 24 \ra^3 \la 6 \ve 5+7 \ve 1] \la 7 \ve 6+5 \ve 1]^2
\la 7 \ve (6+5) (2+3) \ve 4 \ra} {s_{234} s_{567} \la 23 \ra \la 34 \ra
\la 56 \ra \la 67 \ra \la 5 \ve 6+7 \ve 1] \la 4 \ve 3+2 \ve 1] 
\la 7 \ve (6+5) (4+3) \ve 2 \ra}\\
&-\frac{[16]^2 \la 24 \ra^3 \la 4 \ve 1+7 \ve 6]}{s_{671} \la 23 \ra 
\la 34 \ra \la 45 \ra [67] \la 2 \ve 1+7 \ve 6] \la 5 \ve 6+7 \ve 1]} \, .
\end{split}
\eeq

Choosing $j=4, l=5$, $A_3$ is given by
\beq
\begin{split}
A_3&= A_7^{(0)}(1_q^+,2_{\bar{q}}^-,3^-,4^+,5^+,6_p^+, 7_{\bar{p}}^-)= \\
&-\frac{\la 27\ra^2 s_{456}^2 \la 6 \ve (5+4) (7+1) \ve 2 \ra}{s_{712}
\la 12 \ra \la 45 \ra \la 56 \ra \la 6 \ve 5+4 \ve 3] \la 7 \ve 1+2 \ve 3]
\la 2 \ve (1+7) (6+5) \ve 4 \ra}\\
&-\frac{\la 6 \ve 3+2 \ve 1] \la 7 \ve 1+3 \ve 2] \la 7 \ve 3+2 \ve 1]^2}
{s_{123} [12] [23] \la 45 \ra \la 56 \ra \la 67 \ra \la 4 \ve 3+2 \ve 1]
\la 7 \ve 1+2 \ve 3]}\\
&+\frac{\la 46 \ra \la 23 \ra^2 \la 4 \ve 5+6 \ve 1]^2 
\la 3 \ve (7+1) (6+5) \ve 4 \ra}{\la 34 \ra \la 45 \ra \la 56 \ra 
\la 4 \ve 5+6 \ve 7] \la 4 \ve 3+2 \ve 1] \la 6 \ve (7+1) (2+3) \ve 4 \ra
\la 2 \ve (1+7) (6+5) \ve 4 \ra}\\
&-\frac{\la 13 \ra \la 23 \ra^2 [57] [56]^2}{s_{567} \la 12 \ra \la 34 \ra
[67] \la 4 \ve 5+6 \ve 7] \la 1 \ve 7+6 \ve 5]}\\
&-\frac{\la 23 \ra^2 \la 3 \ve 2+4 \ve 5] \la 4 \ve 3+2 \ve 5]
\la 6 \ve 7+1 \ve 5] \la 7 \ve 1+6 \ve 5]^2}{s_{671} s_{234} \la 34 \ra 
\la 67 \ra \la 2 \ve 3+4 \ve 5] \la 4 \ve 3+2 \ve 5] \la 1 \ve 7+6 \ve 5]
\la 6 \ve (7+1) (2+3) \ve 4 \ra}\\
&-\frac{\la 26 \ra \la 27 \ra^2 [45]^3}{s_{345} \la 12 \ra [34] \la 67 \ra 
\la 2 \ve 3+4 \ve 5] \la 6 \ve 5+4 \ve 3]} \, .
\end{split}
\eeq

Picking $j=3, l=4$ we have obtained for $A_4$ the following expression
\beq
\begin{split}
A_4&= A_7^{(0)}(1_q^+,2_{\bar{q}}^-,3^+,4^+,5^-,6_p^-, 7_{\bar{p}}^+)= \\
&-\frac{[17]^2 \la 25 \ra^3 \la 5 \ve 1+7 \ve 6] }{s_{671} \la 23 \ra 
\la 34 \ra \la 45 \ra [67] \la 5 \ve 6+7 \ve 1] \la 2 \ve 1+7 \ve 6] } \\
&-\frac{\la 5 \ve 4+3 \ve 6] \la 2 \ve (1+7) (3+4) \ve 5 \ra^3 }
{s_{712} s_{345} \la 12 \ra \la 34 \ra \la 45 \ra \la 3 \ve 4+5 \ve 6]
\la 2 \ve 1+7 \ve 6] \la 7 \ve (1+2) (3+4) \ve 5 \ra} \\
&+\frac{\la 57 \ra \la 56 \ra^2 \la 5 \ve 4+3 \ve 1]^3}
{[12] \la 34 \ra \la 45 \ra \la 67 \ra \la 5 \ve 6+7 \ve 1] 
\la 5 \ve (6+7) (1+2) \ve 3 \ra \la 7 \ve (1+2) (3+4) \ve 5 \ra} \\
&+\frac{[46]\la 2 \ve 6+5 \ve 4]^3}{s_{456} \la 12 \ra \la 23 \ra 
[45] [56] \la 7 \ve 6+5 \ve 4] \la 3 \ve 4+5 \ve 6]} \\
&-\frac{\la 57 \ra \la 56 \ra^2 \la 2 \ve 1+3 \ve 4]^3}
{s_{567} s_{123} \la 12 \ra \la 23 \ra \la 67 \ra \la 7 \ve 6+5 \ve 4] 
\la 5 \ve (6+7) (1+2) \ve 3 \ra }  \, . \\
\end{split}
\eeq

$A_5$ has been calculated by taking $j=4, l=5$. This gives
\beq
\begin{split}
A_5&= A_7^{(0)}(1_q^+,2_{\bar{q}}^-,3^+,4^-,5^+,6_p^-, 7_{\bar{p}}^+)= \\
&-\frac{[17]^2 \la 24 \ra^3 \la 46 \ra^3 \la 4 \ve (5+6) (3+2) \ve 4 \ra }
{\la 23 \ra \la 34 \ra \la 45 \ra \la 56 \ra 
\la 4 \ve 5+6 \ve 7] \la 4 \ve 3+2 \ve 1] 
\la 4 \ve (5+6) (7+1) \ve 2 \ra \la 6 \ve (7+1) (2+3) \ve 4 \ra } \\
&-\frac{\la 46 \ra^3 \la 4 \ve 5+6 \ve 3] \la 2 \ve 1+7 \ve 3]^3}{s_{456} 
s_{712} \la 12 \ra \la 45 \ra \la 56 \ra \la 6 \ve 5+4 \ve 3] 
\la 7 \ve 1+2 \ve 3] \la 4 \ve (5+6) (7+1) \ve 2 \ra} \\
&-\frac{\la 47 \ra \la 46 \ra^3 [13]^3}{s_{123} [12] \la 45 \ra \la 56 \ra
\la 67 \ra \la 7 \ve 1+2 \ve 3] \la 4 \ve 3+2 \ve 1]} \\
&-\frac{\la 14 \ra \la 24 \ra^3 [57]^3}{s_{567} \la 12 \ra \la 23 \ra 
\la 34 \ra [67] \la 4 \ve 5+6 \ve 7] \la 1 \ve 7+6 \ve 5]} \\
&-\frac{\la 24 \ra^3 \la 4 \ve 3+2 \ve 5] \la 6 \ve 7+1 \ve 5]^3}
{s_{671} s_{234} \la 23 \ra \la 34 \ra \la 67 \ra \la 2 \ve 3+4 \ve 5] 
\la 1 \ve 7+6 \ve 5] \la 6 \ve (7+1) (2+3) \ve 4 \ra } \\
&-\frac{\la 26 \ra^3 [35]^4}{s_{345} \la 12 \ra [34] [45] \la 67 \ra 
\la 2 \ve 3+4 \ve 5] \la 6 \ve 5+4 \ve 3]} \, . \\
\end{split}
\eeq

$A_6$ was computed by choosing $j=7, l=1$, arriving at
\beq
\begin{split}
A_6&= A_7^{(0)}(1_q^+,2_{\bar{q}}^-,3^+,4^+,5_p^+, 6_{\bar{p}}^-,7^-)= \\
&-\frac{\la 27 \ra^3 \la 5 \ve 6+4 \ve 3] \la 6 \ve 5+4 \ve 3]^2}
{s_{456} s_{712} \la 12 \ra \la 45 \ra \la 56 \ra \la 7 \ve 1+2 \ve 3]
\la 2 \ve (1+7) (6+5) \ve 4 \ra } \\
&+\frac{\la 27 \ra^3 \la 2 \ve 3+4 \ve 6] \la 2 \ve 3+4 \ve 5]^2}
{\la 12 \ra \la 23 \ra \la 34 \ra [56] \la 2 \ve 1+7 \ve 6]
\la 2 \ve (3+4) (5+6) \ve 7 \ra  \la 2 \ve (1+7) (6+5) \ve 4 \ra } \\
&-\frac{\la 57 \ra \la 67 \ra^2 [13]^3}{s_{123} [12] \la 45 \ra \la 56 \ra
\la 4 \ve 3+2 \ve 1] \la 7 \ve 1+2 \ve 3] } \\
&+\frac{\la 57 \ra \la 67 \ra^2 \la 2 \ve 3+4 \ve 1]^3}
{s_{567} \la 23 \ra \la 34 \ra \la 56 \ra \la 5 \ve 6+7 \ve 1] 
\la 4 \ve 3+2 \ve 1] \la 2 \ve (3+4) (5+6) \ve 7 \ra } \\
&-\frac{\la 25 \ra [16] \la 2 \ve 6+7 \ve 1]^2}{\la 23 \ra \la 34 \ra 
\la 45 \ra [67] [71] \la 2 \ve 1+7 \ve 6] \la 5 \ve 6+7 \ve 1]} \, . \\
\end{split}
\eeq

$A_7$ was obtained by picking $j=4, l=5$. The final result gives
\beq
\begin{split}
A_7&= A_7^{(0)}(1_q^+,2_{\bar{q}}^-,3^+,4^-,5_p^+, 6_{\bar{p}}^-,7^+)= \\
&+\frac{[17]^2 \la 24 \ra^3 \la 46 \ra^3 \la 4 \ve 3+2 \ve 7]}
{\la 23 \ra \la 34 \ra \la 56 \ra \la 4 \ve 5+6 \ve 7] \la 4 \ve 3+2 \ve 1]
\la 6 \ve (7+1) (2+3) \ve 4 \ra \la 4 \ve (5+6) (7+1) \ve 2 \ra} \\
&-\frac{\la 46 \ra^3 \la 1 \ve 7+2 \ve 3] \la 2 \ve 1+7 \ve 3]^3}
{s_{456} s_{712} \la 56 \ra \la 71 \ra \la 12 \ra \la 6 \ve 5+4 \ve 3]
\la 7 \ve 1+2 \ve 3] \la 4 \ve (5+6) (7+1) \ve 2 \ra}\\
&-\frac{[13]^3 \la 46 \ra^3}{s_{123} [12] \la 56 \ra \la 67 \ra 
\la 7 \ve 1+2 \ve 3] \la 4 \ve 3+2 \ve 1]}\\
&-\frac{\la 14 \ra \la 24 \ra^3 [57]^3}{s_{567} \la 12 \ra \la 23 \ra 
\la 34 \ra [56] \la 1 \ve 7+6 \ve 5] \la 4 \ve 5+6 \ve 7]} \\
&-\frac{\la 16 \ra \la 24 \ra^3 \la 4 \ve 3+2 \ve 5] \la 6 \ve 7+1 \ve 5]^2}
{s_{234} \la 23 \ra \la 34 \ra \la 67 \ra \la 71 \ra \la 2 \ve 3+4 \ve 5] 
\la 1 \ve 7+6 \ve 5] \la 6 \ve (7+1) (2+3) \ve 4 \ra}\\
&-\frac{\la 16 \ra \la 26 \ra^2 [35]^3 \la 2 \ve 5+4 \ve 3]}
{s_{345} \la 67 \ra \la 71 \ra \la 12 \ra [34] [45] \la 2 \ve 3+4 \ve 5]
\la 6 \ve 5+4 \ve 3]} \, .
\end{split}
\eeq

We have computed $A_8$ by selecting $j=6, l=7$
\beq
\begin{split}
A_8&= A_7^{(0)}(1_q^+,2_{\bar{q}}^-,3^-,4^+,5_p^+, 6_{\bar{p}}^-,7^+)= \\
&-\frac{\la 16 \ra \la 26 \ra^2 [45]^2 \la 2 \ve 3+5 \ve 4]}{s_{345} 
\la 12 \ra [34] \la 67 \ra \la 71 \ra 
\la 2 \ve 3+4 \ve 5] \la 6 \ve 5+4 \ve 3]} \\
&-\frac{\la 16 \ra \la 6 \ve 5+4 \ve 2] \la 6 \ve (7+1) (2+3) \ve 5 \ra
\la 6 \ve (7+1) (2+3) \ve 6 \ra^2}{[23] \la 45 \ra \la 56 \ra \la 67 \ra
\la 71 \ra  \la 6 \ve 7+1 \ve 2]  \la 6 \ve 5+4 \ve 3]
\la 6 \ve (5+4) (3+2) \ve 1 \ra \la 6 \ve (7+1) (2+3) \ve 4 \ra} \\
&-\frac{\la 16 \ra \la 23 \ra^2 \la 3 \ve 2+4 \ve 5] \la 6 \ve 7+1 \ve 5]^2}
{s_{234} \la 34 \ra \la 67 \ra \la 71 \ra \la 1 \ve 7+6 \ve 5] 
\la 2 \ve 3+4 \ve 5] \la 6 \ve (7+1) (2+3) \ve 4 \ra}\\
&+\frac{\la 35 \ra \la 36 \ra^3 [27] [17]^2}{s_{712} [12] \la 34 \ra 
\la 45 \ra \la 56 \ra \la 6 \ve 7+1 \ve 2] \la 3 \ve 2+1 \ve 7]}\\
&+\frac{\la 13 \ra \la 23 \ra^2 \la 5 \ve 4+6 \ve 7] \la 6 \ve 5+4 \ve 7]^3}
{s_{123} s_{456} \la 12 \ra \la 45 \ra \la 56 \ra \la 4 \ve 5+6 \ve 7]
\la 3 \ve 2+1 \ve 7] \la 6 \ve (5+4) (3+2) \ve 1 \ra } \\
&-\frac{\la 13 \ra \la 23 \ra^2 [57]^3}{s_{567} \la 12 \ra \la 34 \ra [56] 
 \la 4 \ve 5+6 \ve 7]  \la 1 \ve 7+6 \ve 5]} \, . \\
\end{split}
\eeq

Choosing $j=2, l=3$, we arrive at the following expression for $A_9$ 
\beq
\begin{split}
A_9&= A_7^{(0)}(1_q^+,2_{\bar{q}}^-,3^+,4^+,5_p^-, 6_{\bar{p}}^+,7^-)= \\
&-\frac{\la 25 \ra^3 [16]^3}{\la 23 \ra \la 34 \ra \la 45 \ra [67] [71]
\la 5 \ve 6+7 \ve 1] \la 2 \ve 1+7 \ve 6]} \\
&-\frac{\la 27 \ra^3  \la 2 \ve 3+4 \ve 6]^3}{\la 12 \ra 
\la 23 \ra \la 34 \ra [56]  \la 2 \ve 1+7 \ve 6] 
\la 4 \ve (5+6) (7+1) \ve 2 \ra \la 2 \ve (3+4) (5+6) \ve 7 \ra } \\
&+\frac{\la 57 \ra^3  \la 2 \ve 3+4 \ve 1]^3}{s_{567} \la 23 \ra \la 34 \ra
\la 56 \ra  \la 4 \ve 3+2 \ve 1]  \la 5 \ve 6+7 \ve 1]
\la 2 \ve (3+4) (5+6) \ve 7 \ra } \\
&+\frac{\la 27 \ra^3 \la 5 \ve 6+4 \ve 3]^3}{s_{456} s_{712} \la 12 \ra 
\la 45 \ra \la 56 \ra \la 7 \ve 1+2 \ve 3] \la 4 \ve (5+6) (7+1) \ve 2 \ra } \\
&-\frac{[13]^3 \la 57 \ra^3}{s_{123} [12] \la 45 \ra \la 56 \ra
\la 7 \ve 1+2 \ve 3] \la 4 \ve 3+2 \ve 1] } \, . \\
\end{split}
\eeq

The computation of $A_{10}$ has been performed by taking $j=2, l=3$, leading to
\beq
\begin{split}
A_{10}&= A_7^{(0)}(1_q^+,2_{\bar{q}}^-,3^+,4^-,5_p^-, 6_{\bar{p}}^+,7^+)= \\
&+\frac{\la 46 \ra \la 45 \ra^2 \la 24 \ra^3 [17]^2 \la 4 \ve 3+2 \ve 7]}
{\la 23 \ra \la 34 \ra \la 56 \ra \la 4 \ve 3+2 \ve 1] \la 4 \ve 5+6 \ve 7]
\la 4 \ve (5+6) (7+1) \ve 2 \ra \la 6 \ve (7+1) (2+3) \ve 4 \ra } \\
&-\frac{\la 14 \ra \la 24 \ra^3 [57] [67]^2}{s_{567} \la 12 \ra \la 23 \ra 
\la 34 \ra [56] \la 4 \ve 5+6 \ve 7] \la 1 \ve 7+6 \ve 5]} \\
&-\frac{(s_{671})^2 \la 16 \ra \la 24 \ra^3 \la 4 \ve 3+2 \ve 5]}{s_{234} 
\la 23 \ra \la 34 \ra \la 67 \ra \la 71 \ra \la 2 \ve 3+4 \ve 5]
\la 1 \ve 7+6 \ve 5] \la 6 \ve (7+1) (2+3) \ve 4 \ra } \\
&-\frac{\la 16\ra [35] \la 2 \ve 5+4 \ve 3]^3}{s_{345} [34] [45] \la 67 \ra
\la 71 \ra \la 12 \ra \la 2 \ve 3+4 \ve 5] \la 6 \ve 5+4 \ve 3]} \\
&-\frac{\la 46 \ra \la 45 \ra^2 \la 1 \ve 7+2 \ve 3] \la 2 \ve 1+7 \ve 3]^3}
{s_{456} s_{712} \la 56 \ra \la 71 \ra \la 12 \ra \la 7 \ve 1+2 \ve 3]
\la 6 \ve 5+4 \ve 3] \la 4 \ve (5+6) (7+1) \ve 2 \ra} \\
&-\frac{\la 46 \ra \la 45\ra^2 [13]^3}{ s_{123} [12] \la 56 \ra \la 67 \ra
\la 4 \ve 3+2 \ve 1] \la 7 \ve 1+2 \ve 3]} \, . \\
\end{split}
\eeq

We have selected $j=3, l=4$ for the calculation of $A_{11}$, arriving at
\beq
\begin{split}
A_{11}&= A_7^{(0)}(1_q^-,2_{\bar{q}}^+,3^+,4^+,5^-,6_p^+, 7_{\bar{p}}^-)= \\
&-\frac{\la 71 \ra^2 \la 5 \ve 4+3 \ve 6]^3 \la 2 \ve (1+7) (3+4) \ve 5 \ra}
{s_{712} s_{345} \la 12 \ra \la 34 \ra \la 45 \ra \la 2 \ve 1+7 \ve 6]
\la 3 \ve 4+5 \ve 6] \la 7 \ve (1+2) (3+4) \ve 5 \ra} \\
&-\frac{\la 25 \ra \la 5 \ve 1+7 \ve 6]^3}{s_{671} \la 23 \ra 
\la 34 \ra \la 45 \ra [67] \la 2 \ve 1+7 \ve 6] \la 5 \ve 6+7 \ve 1]} \\
&+\frac{\la 57 \ra^3 \la 5 \ve 4+3 \ve 1] \la 5 \ve 4+3 \ve 2]^2 }
{[12] \la 34 \ra \la 45 \ra \la 67 \ra \la 5 \ve 6+7 \ve 1]
\la 5 \ve (6+7) (1+2) \ve 3 \ra \la 7 \ve (1+2) (3+4) \ve 5 \ra} \\
&+\frac{\la 71 \ra^2 [46]^3 \la 2 \ve 6+5 \ve 4] }{s_{456} 
\la 12 \ra \la 23 \ra [45] [56] \la 3 \ve 4+5 \ve 6] \la 7 \ve 6+5 \ve 4]} \\
&-\frac{\la 57 \ra^3 \la 2 \ve 1+3 \ve 4] \la 1 \ve 2+3 \ve 4]^2}
{s_{567} s_{123} \la 12 \ra \la 23 \ra \la 67 \ra \la 7 \ve 6+5 \ve 4]
\la 5 \ve (6+7) (1+2) \ve 3 \ra } \, .\\
\end{split}
\eeq

By choosing $j=4, l=5$, we have computed $A_{12}$
\beq
\begin{split}
A_{12}&= A_7^{(0)}(1_q^-,2_{\bar{q}}^+,3^+,4^-,5^+,6_p^+, 7_{\bar{p}}^-)= \\
&-\frac{\la 46 \ra \la 71 \ra^2 \la 2 \ve 1+7 \ve 3] \la 4 \ve 5+6 \ve 3]^3}
{s_{456} s_{712} \la 12 \ra \la 45 \ra \la 56 \ra \la 6 \ve 5+4 \ve 3]
\la 7 \ve 1+2 \ve 3] \la 4 \ve (5+6) (7+1) \ve 2 \ra } \\
&+\frac{\la 24 \ra \la 46 \ra \la 4 \ve (3+2) (6+5) \ve 4 \ra^3} {\la 23 \ra 
\la 34 \ra \la 45 \ra \la 56 \ra  \la 4 \ve 3+2 \ve 1] \la 4 \ve 5+6 \ve 7] 
\la 6 \ve (7+1) (2+3) \ve 4 \ra \la 4 \ve (5+6) (7+1) \ve 2 \ra} \\
&-\frac{\la 46 \ra \la 47 \ra^3 [13] [23]^2} {s_{123} [12] \la 45 \ra
\la 56 \ra \la 67 \ra \la 4 \ve 3+2 \ve 1] \la 7 \ve 1+2 \ve 3]} \\
&-\frac{\la 24 \ra \la 14 \ra^3 [57] [56]^2}{s_{567} \la 12 \ra \la 23 \ra
\la 34 \ra [67] \la 4 \ve 5+6 \ve 7] \la 1 \ve 7+6 \ve 5]} \\
&-\frac{\la 24 \ra \la 71 \ra^2 \la 6 \ve 7+1 \ve 5] \la 4 \ve 3+2 \ve 5]^3 }
{s_{671} s_{234} \la 23 \ra \la 34 \ra \la 67 \ra \la 1 \ve 7+6 \ve 5]
\la 2 \ve 3+4 \ve 5] \la 6 \ve (7+1) (2+3) \ve 4 \ra} \\
&-\frac{\la 26 \ra \la 71 \ra^2 [35]^4}{s_{345} \la 67 \ra \la 12 \ra 
[34] [45] \la 2 \ve 3+4 \ve 5] \la 6 \ve 5+4 \ve 3]} \, .\\
\end{split}
\eeq

Picking $j=4, l=5$, $A_{13}$ is given by
\beq
\begin{split}
A_{13}&= A_7^{(0)}(1_q^-,2_{\bar{q}}^+,3^+,4^+,5_p^+, 6_{\bar{p}}^-,7^-)= \\
&-\frac{(s_{345})^2 \la 27 \ra \la 17 \ra^2 \la 5 \ve 4+3 \ve 6]}{s_{712}
\la 12 \ra \la 34 \ra \la 45 \ra \la 2 \ve 1+7 \ve 6] \la 3 \ve 4+5 \ve 6]
\la 7 \ve (1+2) (3+4) \ve 5 \ra} \\
&+\frac{\la 57 \ra \la 67 \ra^2 \la 5 \ve 4+3 \ve 1] \la 5 \ve 4+3 \ve 2]^2}
{[12] \la 34 \ra \la 45 \ra \la 56 \ra \la 5 \ve 6+7 \ve 1]      
\la 7 \ve (1+2) (3+4) \ve 5 \ra \la 5 \ve (6+7) (1+2) \ve 3 \ra} \\
&-\frac{s_{671}^2 \la 25 \ra [16]}{\la 23 \ra \la 34 \ra \la 45 \ra 
[67] [71] \la 2 \ve 1+7 \ve 6] \la 5 \ve 6+7 \ve 1]} \\
&+\frac{\la 27 \ra \la 71 \ra^2 [46] [45]^2}{s_{456} \la 12 \ra \la 23 \ra 
[56] \la 7 \ve 6+5 \ve 4] \la 3 \ve 4+5 \ve 6]} \\
&-\frac{\la 57 \ra \la 67 \ra^2 \la 2 \ve 1+3 \ve 4] \la 1 \ve 2+3 \ve 4]^2}
{s_{567} s_{123} \la 12 \ra \la 23 \ra \la 56 \ra \la 7 \ve 6+5 \ve 4]  
\la 5 \ve (6+7) (1+2) \ve 3 \ra} \, . \\
\end{split}
\eeq

Finally, $A_{14}$ has been computed by selecting $j=1, l=2$
\beq
\begin{split}
A_{14}&= A_7^{(0)}(1_q^-,2_{\bar{q}}^+,3^+,4^-,5_p^+, 6_{\bar{p}}^-,7^+)= \\
&-\frac{\la 46 \ra^3 \la 2 \ve 1+7 \ve 3] \la 1 \ve 7+2 \ve 3]^3}
{s_{456} s_{712} \la 56 \ra \la 71 \ra \la 12 \ra \la 6 \ve 5+4 \ve 3]
\la 7 \ve 1+2 \ve 3] \la 4 \ve (5+6) (7+1) \ve 2 \ra} \\
&-\frac{[13] [23]^2 \la 46 \ra^3}{s_{123} [12] \la 56 \ra \la 67 \ra
 \la 4 \ve 3+2 \ve 1] \la 7 \ve 1+2 \ve 3]} \\
&+\frac{\la 24 \ra \la 46 \ra^3 \la 4 \ve 3+2 \ve 7]^3}{\la 23 \ra
\la 34 \ra \la 56 \ra \la 4 \ve 3+2 \ve 1] \la 4 \ve 5+6 \ve 7]
\la 6 \ve (7+1) (2+3) \ve 4 \ra \la 4 \ve (5+6) (7+1) \ve 2 \ra} \\
&-\frac{\la 24 \ra \la 14 \ra^3 [57]^3}{s_{567} \la 12 \ra 
\la 23 \ra \la 34 \ra [56]  \la 1 \ve 7+6 \ve 5]  \la 4 \ve 5+6 \ve 7]} \\
&-\frac{\la 24 \ra \la 16 \ra^3  \la 4 \ve 3+2 \ve 5]^3}{s_{234}
\la 23 \ra \la 34 \ra \la 67 \ra \la 71 \ra  \la 2 \ve 3+4 \ve 5]
 \la 1 \ve 7+6 \ve 5]  \la 6 \ve (7+1) (2+3) \ve 4 \ra } \\
&-\frac{ \la 16 \ra^3 [35]^3 \la 2 \ve 5+4 \ve 3]}
{s_{345} [34] [45] \la 67 \ra \la 71 \ra \la 12 \ra 
\la 6 \ve 5+4 \ve 3] \la 2 \ve 3+4 \ve 5]} \, . 
\end{split}
\eeq
\subsection{4q+3g sub-leading amplitudes}
We anticipate that the sub-leading amplitudes have fewer and simpler terms than the 
leading ones,mainly due to the fact that the ordering of the quarks causes the vanishing of 
many BCFW sub-amplitudes.

The calculation of $B_1$ was performed by taking $j=5, l=6$
\beq
\begin{split}
B_1&= A_7^{(0)}(1_q^+,2_{\bar{p}}^-,3^+,4^+,5^-,6_p^+, 7_{\bar{q}}^-)= \\
&+\frac{[16]^2 \la 25 \ra^3}{s_{671} \la 23 \ra \la 34 \ra \la 45 \ra [71] 
\la 2 \ve 1+7 \ve 6]} \\
&+\frac{\la 27 \ra^2 \la 5 \ve 4+3 \ve 6]^3}{s_{712} s_{345} 
\la 34 \ra \la 45 \ra \la 71 \ra \la 3 \ve 4+5 \ve 6] \la 2 \ve 1+7 \ve 6]}\\
&-\frac{ \la 27 \ra^2 [46]^3}{s_{456} \la 23 \ra [45] [56] \la 71 \ra 
\la 3 \ve 4+5 \ve 6] } \, .\\
\end{split}
\eeq

$B_2$ was computed by selecting $j=2, l=3$. The final result is
\beq
\begin{split}
B_2&= A_7^{(0)}(1_q^+,2_{\bar{p}}^-,3^+,4^-,5^+,6_p^+, 7_{\bar{q}}^-)= \\
&+\frac{\la 46 \ra \la 24 \ra^3 \la 4 \ve 5+6 \ve 1]^2}
{\la 23 \ra \la 34 \ra \la 45 \ra \la 56 \ra [71]
\la 6 \ve (7+1) (2+3) \ve 4 \ra \la 4 \ve (5+6) (7+1) \ve 2 \ra} \\
&+\frac{\la 46 \ra \la 27 \ra^2 \la 4 \ve 5+6 \ve 3]^3}{s_{456} s_{712}
\la 45 \ra \la 56 \ra \la 71 \ra \la 6 \ve 5+4 \ve 3]
\la 4 \ve (5+6) (7+1) \ve 2 \ra} \\
&+\frac{\la 24 \ra^3 \la 4 \ve 3+2 \ve 5] \la 7 \ve 1+6 \ve 5]^2}
{s_{671} s_{234} \la 23 \ra \la 34 \ra \la 71 \ra \la 2 \ve 3+4 \ve 5]
\la 6 \ve (7+1) (2+3) \ve 4 \ra} \\
&-\frac{\la 27 \ra^2 [35]^4}{s_{345} [34] [45] \la  71 \ra 
\la 2 \ve 3+4 \ve 5] \la 6 \ve 5+4 \ve 3]} \, . \\
\end{split}
\eeq

$B_3$ has been obtained by choosing $j=3, l=4$ 
\beq
\begin{split}
B_3&= A_7^{(0)}(1_q^+,2_{\bar{p}}^-,3^-,4^+,5^+,6_p^+, 7_{\bar{q}}^-)= \\
&+\frac{\la 36 \ra \la 3 \ve 7+2 \ve 1]^2}{s_{712} \la 34 \ra 
\la 45 \ra \la 56 \ra [71] \la 6 \ve 7+1 \ve 2]} \\
&+\frac{\la 3 \ve 4+5 \ve 2] \la 3 \ve (4+5) (6+1) \ve 7 \ra^2}
{s_{345} s_{671} \la 34 \ra \la 45 \ra \la 71 \ra 
\la 5 \ve 4+3 \ve 2] \la 6 \ve 7+1 \ve 2] } \\
&-\frac{[24] \la 7 \ve 2+3 \ve 4]^2}{s_{234} [23] [34] 
\la 56 \ra \la 71 \ra \la 5 \ve 4+3 \ve 2]} \, . \\
\end{split}
\eeq

The calculation of $B_4$ has been performed by selecting $j=3, l=4$
\beq
\begin{split}
B_4&= A_7^{(0)}(1_q^+,2_{\bar{p}}^+,3^+,4^+,5^-,6_p^-, 7_{\bar{q}}^-)= \\
&+\frac{\la 25 \ra \la 5 \ve 6+7 \ve 1]^2}{s_{671} \la 23 \ra 
\la 34 \ra \la 45 \ra [71] \la 2 \ve 1+7 \ve 6]} \\
&+\frac{\la 5 \ve 4+3 \ve 6] \la 7 \ve (1+2) (3+4) \ve 5 \ra^2}
{s_{712} s_{345} \la 34 \ra \la 45 \ra \la 71 \ra
\la 3 \ve 4+5 \ve 6] \la 2 \ve 1+7 \ve 6] } \\
&-\frac{[46] \la 7 \ve 6+5 \ve 4]^2}{s_{456} \la 23 \ra 
[45] [56] \la 71 \ra \la 3 \ve 4+5 \ve 6]} \, . \\
\end{split}
\eeq

Setting $j=4, l=5$, we have computed $B_5$, arriving at the following expression
\beq
\begin{split}
B_5&= A_7^{(0)}(1_q^+,2_{\bar{p}}^+,3^+,4^-,5^+,6_p^-, 7_{\bar{q}}^-)= \\
&+\frac{\la 24 \ra \la 46 \ra^3 \la 4 \ve 3+2 \ve 1]^2}{\la 23 \ra 
\la 34 \ra \la 45 \ra \la 56 \ra [71] \la 4 \ve (5+6) (7+1) \ve 2 \ra
\la 6 \ve (7+1) (2+3) \ve 4 \ra} \\
&+\frac{\la 46 \ra^3 \la 4 \ve 5+6 \ve 3] \la 7 \ve 1+2 \ve 3]^2}
{s_{456} s_{712} \la 45 \ra \la 56 \ra \la 71 \ra \la 6 \ve 5+4 \ve 3]
\la 4 \ve (5+6) (7+1) \ve 2 \ra} \\
&+\frac{\la 24 \ra \la 67 \ra^2 \la 4 \ve 3+2 \ve 5]^3 }
{s_{671} s_{234} \la 23 \ra \la 34 \ra \la 71 \ra \la 2 \ve 3+4 \ve 5] 
\la 6 \ve (7+1) (2+3) \ve 4 \ra} \\
&-\frac{\la 67 \ra^2 [35]^4}{s_{345} [34] [45] \la 71 \ra 
\la 2 \ve 3+4 \ve 5] \la 6 \ve 5+4 \ve 3]} \, .\\
\end{split}
\eeq

We have calculated $B_6$ by choosing $j=3, l=4$
\beq
\begin{split}
B_6&= A_7^{(0)}(1_q^+,2_{\bar{p}}^+,3^-,4^+,5^+,6_p^-, 7_{\bar{q}}^-)= \\
&+\frac{ [12]^2 \la 36 \ra^3 }{s_{712} \la 34 \ra \la 45 \ra 
\la 56 \ra [71]  \la 6 \ve 7+1 \ve 2]} \\
&+\frac{\la 67 \ra^2 \la 3 \ve 4+5 \ve 2]^3}{s_{345} s_{671} \la 34 \ra 
\la 45 \ra \la 71 \ra  \la 5 \ve 4+3 \ve 2] \la 6 \ve 7+1 \ve 2] } \\
&-\frac{\la 67 \ra^2 [24]^3}{s_{234} [23] [34] \la 56 \ra \la 71\ra
\la 5 \ve 4+3 \ve 2 } \, .\\
\end{split}
\eeq

Picking $j=4, l=5$, $B_7$ is simply given by
\beq
\begin{split}
B_7&= A_7^{(0)}(1_q^+,2_{\bar{p}}^-,3^+,4^+,5_p^+, 6_{\bar{q}}^-,7^-)= \\
&-\frac{ \la 2 \ve 6+7 \ve 1]^2}
{s_{671} \la 23 \ra\la 34 \ra \la 45 \ra [67] [71]} \, . \\
\end{split}
\eeq

Selecting $j=4, l=5$, the calculation of $B_8$ yields
\beq
\begin{split}
B_8&= A_7^{(0)}(1_q^+,2_{\bar{p}}^-,3^+,4^-,5_p^+, 6_{\bar{q}}^-,7^+)= \\
&+\frac{\la 24 \ra^3 \la 6 \ve 7+1 \ve 5]^2}{s_{671} s_{234}
\la 23 \ra \la 34 \ra \la 67 \ra \la 71 \ra \la 2 \ve 3+4 \ve 5]} \\
&-\frac{\la 26 \ra^2 [35]^3}{s_{345} [34] [45] 
\la 67 \ra \la 71 \ra \la 2 \ve 3+4 \ve 5]} \, . \\
\end{split}
\eeq

The last four sub-leading amplitudes $B_9 - B_{12}$ were obtained by choosing $j=6, l=7$, and read 
\beq
\begin{split}
B_9&= A_7^{(0)}(1_q^+,2_{\bar{p}}^-,3^-,4^+,5_p^+, 6_{\bar{q}}^-,7^+)= \\
&+\frac{\la 35 \ra \la 6 \ve (7+1) (5+4) \ve 3 \ra^2}{s_{345} s_{671} 
\la 34 \ra \la 45 \ra \la 67 \ra \la 71 \ra \la 5 \ve 4+3 \ve 2]} \\
&-\frac{[24] \la 6 \ve 2+3 \ve 4]^2}{s_{234} [23] [34] 
\la 67 \ra \la 71 \ra \la 5 \ve 4+3 \ve 2]} \, ,\\
\end{split}
\eeq

\beq
\begin{split}
B_{10}&= A_7^{(0)}(1_q^+,2_{\bar{p}}^+,3^+,4^+,5_p^-, 6_{\bar{q}}^-,7^-)= \\
&-\frac{\la 5 \ve 6+7 \ve 1]^2}
{s_{671} \la 23 \ra \la 34 \ra \la 45 \ra [67] [71]} \, .
\end{split}
\eeq

\beq
\begin{split}
B_{11}&= A_7^{(0)}(1_q^+,2_{\bar{p}}^+,3^+,4^-,5_p^-, 6_{\bar{q}}^-,7^+)= \\
&+\frac{\la 24 \ra \la 6 \ve (7+1) (2+3) \ve 4 \ra^2}{s_{671} s_{234}
\la 23 \ra \la 34 \ra \la 67 \ra \la 71 \ra \la 2 \ve 3+4 \ve 5]} \\
&-\frac{[35] \la 6 \ve 5+4 \ve 3]^2}{s_{345} [34] [45] 
\la 67 \ra \la 71 \ra \la 2 \ve 3+4 \ve 5]} \, ,
\end{split}
\eeq

\beq
\begin{split}
B_{12}&= A_7^{(0)}(1_q^+,2_{\bar{p}}^+,3^-,4^+,5_p^-, 6_{\bar{q}}^-,7^+)= \\
&+\frac{\la 35 \ra^3 \la 6 \ve 7+1 \ve 2]^2}{s_{345} s_{671}
\la 34 \ra \la 45 \ra \la 67 \ra \la 71 \ra \la 5 \ve 4+3 \ve 2]} \\
&-\frac{\la 56 \ra^2 [24]^3}{s_{234} [23] [34] \la 67 \ra \la 71 \ra
\la 5 \ve 4+3 \ve 2]} \, . \\
\end{split}
\eeq

\section{NMHV six-quark plus one-gluon amplitudes}
Six-quark processes in QCD were first considered in \cite{Gunion:1986zh}. We discuss here
the three different flavor ($p$, $q$ and $r$) case. In order to obtain the amplitudes for two or 
three equal flavors one has to antisymmetrize the expression by adding and substracting  
amplitudes with the antiquarks exchanged; for more details see \cite{Gunion:1986zh}.
From Eq.(\ref{qgcolor}), one can notice that three different color structure arise: the leading, 
sub-leading and sub-sub-leading contributions, corresponding to $p=0$, $p=1$, and $p=2$, 
respectively.

For $p=0$, formally the two following possibilities appear : 
$\alpha_1=\bar{i}_2, \alpha_2=\bar{i}_3, \alpha_3=\bar{i}_1 $ and 
$\alpha_1=\bar{i}_3, \alpha_2=\bar{i}_1, \alpha_3=\bar{i}_2 $. The corresponding color 
factors are
\beq
(T^{a(\sigma_1)} \dots T^{a(\sigma_j)})_{i_1 \bar{i}_2} 
(T^{a(\sigma_{j+1})} \dots T^{a(\sigma_l)})_{i_2 \bar{i}_3} 
(T^{a(\sigma_{l+1})} \dots T^{a(\sigma_k)})_{i_3 \bar{i}_1} 
\eeq
and
\beq
(T^{a(\sigma_1)} \dots T^{a(\sigma_j)})_{i_1 \bar{i}_3} 
(T^{a(\sigma_{j+1})} \dots T^{a(\sigma_l)})_{i_2 \bar{i}_1} 
(T^{a(\sigma_{l+1})} \dots T^{a(\sigma_k)})_{i_3 \bar{i}_2} \, ,
\eeq
even though the second structure does not contribute to the physical process in interest.
To complete the color basis, one has to add also those structures corresponding to the  
$2 \leftrightarrow 3$ exchange, but those terms do not lead to independent amplitudes.

For $p=1$, the three following color structures have to be taken into account:
$\alpha_1=\bar{i}_1, \alpha_2=\bar{i}_3, \alpha_3=\bar{i}_2 $,
$\alpha_1=\bar{i}_2, \alpha_2=\bar{i}_1, \alpha_3=\bar{i}_3 $ and 
$\alpha_1=\bar{i}_3, \alpha_2=\bar{i}_2, \alpha_3=\bar{i}_1 $.
It is easy to see
that the second structure can be related to the first one by cyclic permutation, and similarly for 
the third-one. Therefore the only independent sub-leading color form factor corresponds to the 
string
\beq
\frac{-1}{N} (T^{a(\sigma_1)} \dots T^{a(\sigma_j)})_{i_1 \bar{i}_1} 
(T^{a(\sigma_{j+1})} \dots T^{a(\sigma_l)})_{i_2 \bar{i}_3} 
(T^{a(\sigma_{l+1})} \dots T^{a(\sigma_k)})_{i_3 \bar{i}_2} \, .
\eeq
When $p=2$ things become simpler than in the previous cases, because the only possible 
configuration is $\alpha \equiv i$. Then the sub-sub-leading color structure is given by
\beq
\frac{1}{N^2} (T^{a(\sigma_1)} \dots T^{a(\sigma_j)})_{i_1 \bar{i}_1} 
(T^{a(\sigma_{j+1})} \dots T^{a(\sigma_l)})_{i_2 \bar{i}_2} 
(T^{a(\sigma_{l+1})} \dots T^{a(\sigma_k)})_{i_3 \bar{i}_3} \, .
\eeq 
At the level of the partial amplitudes, the string for leading color is written in a generic form 
as
\beq
[q (g_{\sigma_1} \dots g_{\sigma_j}) \bar{p}] 
[p (g_{\sigma_{j+1}} \dots g_{\sigma_l}) \bar{r}]
[r (g_{\sigma_{l+1}} \dots g_{\sigma_l}) \bar{q}]
\eeq
for first leading-color structure and
\beq
[q (g_{\sigma_1} \dots g_{\sigma_j}) \bar{r}] 
[p (g_{\sigma_{j+1}} \dots g_{\sigma_l}) \bar{q}]
[r (g_{\sigma_{l+1}} \dots g_{\sigma_l}) \bar{p}]
\eeq
for the second one.
For the sub-leading color one has
\beq
[q (g_{\sigma_1} \dots g_{\sigma_j}) \bar{q}] 
[p (g_{\sigma_{j+1}} \dots g_{\sigma_l}) \bar{r}]
[r (g_{\sigma_{l+1}} \dots g_{\sigma_l}) \bar{p}] \, ,
\eeq 
and for the sub-sub-leading case the structure is
\beq
[q (g_{\sigma_1} \dots g_{\sigma_j}) \bar{q}] 
[p (g_{\sigma_{j+1}} \dots g_{\sigma_l}) \bar{p}]
[r (g_{\sigma_{l+1}} \dots g_{\sigma_l}) \bar{r}]
\eeq
In order to maintain the same notation along this paper we present the results for the amplitudes 
using the following order of partons ($q$ always first)
\beq
A(q, \bar{q}, g_1, \dots, g_j, p, \bar{p}, g_{j+1}, \dots, g_k, r, \bar{r},
g_{k+1}, \dots g_l) 
\eeq
or
\beq
A(q, \bar{p}, g_1, \dots, g_j, r, \bar{q}, g_{j+1}, \dots, g_k, p, \bar{r},
g_{k+1}, \dots, g_l) 
\eeq
for the leading amplitudes, 
\beq
A(q, \bar{p}, g_1, \dots, g_j, p, \bar{q}, g_{j+1}, \dots, g_k, r, \bar{r},
g_{k+1}, \dots, g_l) 
\eeq
for the sub-leading ones and 
\beq
A(q, \bar{p}, g_1, \dots, g_j, p, \bar{r}, g_{j+1}, \dots, g_k, r, \bar{q},
g_{k+1}, \dots, g_l) 
\eeq
for the sub-sub-leading case.

For the sake of presentation, we denote the leading contributions as 
``A'' amplitudes, the sub-leading as ``B'' and the sub-sub-leading ``C''. We have 
found that the number of independent amplitudes is in principle twenty-four: eight for the 
leading color ($A_1,\dots,A_8$), twelve for the sub-leading 
($B_1,\dots,B_{12}$) and four for the sub-sub-leading ($C_1,\dots,C_{4}$).

\subsection{Leading amplitudes}
In order to obtain a leading amplitude the only available gluon should sit in between
two fermions of different flavor. It is therefore
equivalent, just by flavor relabelling, to insert the gluon in any possible place
 compatible with the color structure. 
We have chosen the gluon to be the third particle. 
Besides, as there exist already three fermions with negative polarization, the gluon should 
necessarily have positive helicity to build an NMHV amplitude. Those coming from the first 
color structure are
\beq
\begin{split}
A_1=A_7^{(0)}(1_q^+,2_{\bar{q}}^-,3^+,4_p^+,5_{\bar{p}}^-,6_r^+, 7_{\bar{r}}^-) \qquad \qquad
&A_3=A_7^{(0)}(1_q^+,2_{\bar{q}}^-,3^+,4_p^-,5_{\bar{p}}^+,6_r^-, 7_{\bar{r}}^+) \\
A_2=A_7^{(0)}(1_q^+,2_{\bar{q}}^-,3^+,4_p^+,5_{\bar{p}}^-,6_r^-, 7_{\bar{r}}^+) \qquad \qquad
&A_4=A_7^{(0)}(1_q^-,2_{\bar{q}}^+,3^+,4_p^+,5_{\bar{p}}^-,6_r^-, 7_{\bar{r}}^+) \, ,\\
\end{split}
\eeq
while those from the second color string 
\beq
\begin{split}
A_5=A_7^{(0)}(1_q^+,2_{\bar{p}}^-,3^+,4_r^+,5_{\bar{q}}^-,6_p^+, 7_{\bar{r}}^-) \qquad \qquad 
&A_7=A_7^{(0)}(1_q^+,2_{\bar{p}}^+,3^+,4_r^+,5_{\bar{q}}^-,6_p^-, 7_{\bar{r}}^-) \\
A_6=A_7^{(0)}(1_q^+,2_{\bar{p}}^-,3^+,4_r^-,5_{\bar{q}}^-,6_p^+, 7_{\bar{r}}^+) \qquad \qquad
&A_8=A_7^{(0)}(1_q^+,2_{\bar{p}}^+,3^+,4_r^-,5_{\bar{q}}^-,6_p^-, 7_{\bar{r}}^+) \\
\end{split}
\eeq
are all vanishing. Below we present the most compact expressions for the four leading amplitudes.

$A_1$ has been obtained by picking  $j=5, l=6$
\beq
\begin{split}
A_1&=A_7^{(0)}(1_q^+,2_{\bar{q}}^-,3^+,4_p^+,5_{\bar{p}}^-,6_r^+, 7_{\bar{r}}^-)= \\
&-\frac{\la 57 \ra^2 [34]^2 \la 27 \ra^3 \la 7 \ve 6+5 \ve 3] }{\la 12 \ra 
\la 67 \ra \la 7 \ve 1+2 \ve 3] \la 7 \ve 6+5 \ve 4] 
\la 2 \ve (3+4) (5+6) \ve 7 \ra \la 7 \ve (1+2) (3+4) \ve 5 \ra } \\
&-\frac{\la 47 \ra \la 57 \ra^2 [13]^3}{s_{123} [12] \la 45\ra \la 67 \ra
\la 4 \ve 3+2 \ve 1] \la 7 \ve 1+2 \ve 3]} \\
&+\frac{\la 24 \ra \la 57 \ra^2 \la 7 \ve 6+5 \ve 1] \la 2 \ve 3+4 \ve 1]^2}
{s_{567} \la 23 \ra \la 34 \ra \la 67 \ra \la 5 \ve 6+7 \ve 1] 
\la 4 \ve 3+2 \ve 1] \la 2 \ve (3+4) (5+6) \ve 7 \ra } \\
&-\frac{\la 24 \ra \la 25 \ra^2 [16]^2 \la 5 \ve 1+7 \ve 6]}{s_{671} \la 23 \ra 
\la 34 \ra \la 45 \ra [67] \la 2 \ve 1+7 \ve 6] \la 5 \ve 6+7 \ve 1]} \\
&-\frac{\la 27 \ra^2 \la 4 \ve 3+5 \ve 6] \la 5 \ve 4+3 \ve 6]^2
 \la 2 \ve (1+7) (3+4) \ve 5 \ra}{s_{712} s_{345} \la 12 \ra \la 34 \ra \la 45 \ra 
\la 3 \ve 4+5 \ve 6]  \la 2 \ve 1+7 \ve 6] \la 7 \ve (1+2) (3+4) \ve 5 \ra } \\
&-\frac{\la 27 \ra^2 [46]^2  \la 2 \ve 6+5 \ve 4]}{s_{456} 
\la 12 \ra \la 23 \ra [45]  \la 3 \ve 4+5 \ve 6]  \la 7 \ve 6+5 \ve 4]} \, . \\
\end{split}
\eeq

Selecting $j=2, l=3$, $A_2$ yields
\beq
\begin{split}
A_2&=A_7^{(0)}(1_q^+,2_{\bar{q}}^-,3^+,4_p^+,5_{\bar{p}}^-,6_r^-, 7_{\bar{r}}^+)= \\
&-\frac{\la 24 \ra \la 25 \ra^2 [71]^2 \la 5 \ve 1+7 \ve 6]}{s_{671} 
\la 23 \ra \la 34 \ra \la 45 \ra [67] \la 5 \ve 6+7 \ve 1] \la 2 \ve 1+7 \ve 6]} \\
&+\frac{\la 24 \ra \la 2 \ve (3+4) (5+6) \ve 2 \ra^3}
{\la 12 \ra \la 23 \ra \la 34 \ra \la 2 \ve 3+4 \ve 5] \la 2 \ve 1+7 \ve 6]
\la 4 \ve (5+6) (7+1) \ve 2 \ra \la 2 \ve (3+4) (5+6) \ve 7 \ra} \\
&+\frac{\la 24 \ra \la 56 \ra^2 \la 7 \ve 6+5 \ve 1] \la 2 \ve 3+4 \ve 1]^2}
{ s_{567} \la 23 \ra \la 34 \ra \la 67 \ra \la 4 \ve 3+2 \ve 1] 
\la 5 \ve 6+7 \ve 1] \la 2 \ve (3+4) (5+6) \ve 7 \ra} \\
&+\frac{[35] [34]^2 \la 26 \ra^3}{s_{345} \la 67 \ra \la 12 \ra [45]
\la 6 \ve 5+4 \ve 3] \la 2 \ve 3+4 \ve 5]} \\
&-\frac{\la 56 \ra^2 \la 4 \ve 5+6 \ve 3] \la 2 \ve 1+7 \ve 3]^3}
{s_{456} s_{712} \la 12 \ra \la 45 \ra \la 7 \ve 1+2 \ve 3] 
\la 6 \ve 5+4 \ve 3] \la 4 \ve (5+6) (7+1) \ve 2 } \\
&-\frac{\la 47 \ra \la 56 \ra^2 [13]^3}{s_{123} [12] 
\la 45 \ra \la 67 \ra \la 7 \ve 1+2 \ve 3] \la 4 \ve 3+2 \ve 1]} \, .\\
\end{split}
\eeq

Picking $j=2, l=3$, $A_3$ is given by
\beq
\begin{split}
A_3&=A_7^{(0)}(1_q^+,2_{\bar{q}}^-,3^+,4_p^-,5_{\bar{p}}^+,6_r^-, 7_{\bar{r}}^+)= \\
&-\frac{\la 46 \ra^2 [71]^2 \la 24 \ra^3 \la 4 \ve (5+6) (2+3) \ve 4 \ra}
{\la 23 \ra \la 34 \ra \la 45 \ra \la 4 \ve 5+6 \ve 7] \la 4 \ve 3+2 \ve 1]
\la 6 \ve (7+1) (2+3) \ve 4 \ra \la 4 \ve (5+6) (7+1) \ve 2 \ra} \\
&+\frac{\la 24 \ra^3 \la 6 \ve 7+1 \ve 5]^3}
{s_{671} \la 23 \ra \la 34 \ra \la 67 \ra 
\la 2 \ve 3+4 \ve 5] \la 1 \ve 7+6 \ve 5]   \la 6 \ve (7+1) (2+3) \ve 4 \ra} \\
&-\frac{[57]^2 \la 24 \ra^3 \la 1 \ve 5+6 \ve 7]}{s_{567} \la 12 \ra 
\la 23 \ra \la 34 \ra [67] \la 4 \ve 5+6 \ve 7] \la 1 \ve 7+6 \ve 5]} \\
&+\frac{ \la 26 \ra^3 [35]^3}{s_{345} \la 67 \ra \la 12 \ra [45]
\la 2 \ve 3+4 \ve 5] \la 6 \ve 5+4 \ve 3]} \\
&-\frac{\la 46 \ra^2 \la 4 \ve 5+6 \ve 3] \la 2 \ve 1+7 \ve 3]^3}
{s_{456} s_{712} \la 12 \ra \la 45 \ra \la 7 \ve 1+2 \ve 3]
\la 6 \ve 5+4 \ve 3] \la 4 \ve (5+6) (7+1) \ve 2 \ra } \\
&-\frac{\la 47 \ra \la 46 \ra^2 [13]^3}{s_{123} 
[12] \la 45 \ra \la 67 \ra \la 7 \ve 1+2 \ve 3] \la 4 \ve 3+2 \ve 1]} \, . \\
\end{split}
\eeq

$A_4$ has been computed by choosing $j=1, l=7$, as
\beq
\begin{split}
A_4&=A_7^{(0)}(1_q^-,2_{\bar{q}}^+,3^+,4_p^+,5_{\bar{p}}^-,6_r^-, 7_{\bar{r}}^+)= \\
&+\frac{ \la 26 \ra \la 16 \ra^2 [35] [34]^2}
{s_{345} \la 67 \ra \la 12 \ra [45] \la 6 \ve 5+4 \ve 3] \la 2 \ve 3+4 \ve 5]} \\
&+\frac{\la 46 \ra \la 56 \ra^2 \la 16 \ra^2 [23]^2 \la 6 \ve 7+1 \ve 3]}
{\la 45 \ra \la 67 \ra \la 6 \ve 7+1 \ve 2] \la 6 \ve 5+4 \ve 3]
\la 1 \ve (2+3) (4+5) \ve 6 \ra \la 6 \ve (7+1) (2+3) \ve 4 \ra } \\
&+\frac{(s_{234})^2 \la 24 \ra \la 16 \ra^2 \la 6 \ve 7+1 \ve 5]}
{s_{671} \la 23 \ra \la 34 \ra \la 67 \ra \la 1 \ve 7+6 \ve 5]
\la 2 \ve 3+4 \ve 5] \la 6 \ve (7+1) (2+3) \ve 4 \ra } \\
&-\frac{ \la 24 \ra \la 1 \ve 5+6 \ve 7]^3}{s_{567} \la 12 \ra 
\la 23 \ra \la 34 \ra [67] \la 4 \ve 5+6 \ve 7] \la 1 \ve 7+6 \ve 5]} \\
&-\frac{\la 56 \ra^2 \la 2 \ve 3+1 \ve 7] \la 1 \ve 2+3 \ve 7]^2
\la 1 \ve (2+3) (6+5) \ve 4 \ra} {s_{123} s_{456} \la 12 \ra \la 23 \ra 
\la 45 \ra  \la 3 \ve 2+1 \ve 7] \la 4 \ve 5+6 \ve 7] 
\la 1 \ve (2+3) (4+5) \ve 6 \ra} \\
&-\frac{\la 56 \ra^2 [72]^2 \la 4 \ve 1+7 \ve 2]}
{s_{712} [12] \la 34 \ra 
\la 45 \ra \la 6 \ve 7+1 \ve 2] \la 3 \ve 2+1 \ve 7]} \, . \\
\end{split}
\eeq
\subsection{Sub-leading amplitudes}
At this level, the gluon may be at the third, fifth or seventh place. 
It is easy to show that two inequivalent choices for NMHV amplitudes are 
$3^+$ and $5^+$. If the gluon is the third particle, one has the 
following amplitudes
\beq
\begin{split}
B_1=A_7^{(0)}(1_q^+,2_{\bar{p}}^-,3^+,4_p^+,5_{\bar{q}}^-,6_r^+, 7_{\bar{r}}^-) \qquad \qquad
&B_3=A_7^{(0)}(1_q^+,2_{\bar{p}}^+,3^+,4_p^-,5_{\bar{q}}^-,6_r^+, 7_{\bar{r}}^-) \\
B_2=A_7^{(0)}(1_q^+,2_{\bar{p}}^-,3^+,4_p^+,5_{\bar{q}}^-,6_r^-, 7_{\bar{r}}^+) \qquad \qquad
&B_4=A_7^{(0)}(1_q^+,2_{\bar{p}}^+,3^+,4_p^-,5_{\bar{q}}^-,6_r^-, 7_{\bar{r}}^+)  \, ,
\end{split}
\eeq
while if the gluon is $5^+$ one has to consider
\beq
\begin{split}
B_5=A_7^{(0)}(1_q^+,2_{\bar{p}}^-,3_p^+,4_{\bar{q}}^-,5^+,6_r^+, 7_{\bar{r}}^-) \qquad \qquad
&B_9=A_7^{(0)}(1_q^-,2_{\bar{p}}^-,3_p^+,4_{\bar{q}}^+,5^+,6_r^+, 7_{\bar{r}}^-) \\
B_6=A_7^{(0)}(1_q^+,2_{\bar{p}}^-,3_p^+,4_{\bar{q}}^-,5^+,6_r^-, 7_{\bar{r}}^+) \qquad \qquad
&B_{10}=A_7^{(0)}(1_q^-,2_{\bar{p}}^-,3_p^+,4_{\bar{q}}^+,5^+,6_r^-, 7_{\bar{r}}^+) \\
B_7=A_7^{(0)}(1_q^+,2_{\bar{p}}^+,3_p^-,4_{\bar{q}}^-,5^+,6_r^+, 7_{\bar{r}}^-) \qquad \qquad
&B_{11}=A_7^{(0)}(1_q^-,2_{\bar{p}}^+,3_p^-,4_{\bar{q}}^+,5^+,6_r^+, 7_{\bar{r}}^-) \\
B_8=A_7^{(0)}(1_q^+,2_{\bar{p}}^+,3_p^-,4_{\bar{q}}^-,5^+,6_r^-, 7_{\bar{r}}^+) \qquad \qquad
&B_{12}=A_7^{(0)}(1_q^-,2_{\bar{p}}^+,3_p^-,4_{\bar{q}}^+,5^+,6_r^-, 7_{\bar{r}}^+)  \, .
\end{split}
\eeq
Half of those amplitudes can be related to the others by a simple exchange of quarks, 
corresponding to a ``flip'' operation. 

By selecting $j=2, l=3$, one has for $B_1$ the following expression
\beq
\begin{split}
B_1&=A_7^{(0)}(1_q^+,2_{\bar{p}}^-,3^+,4_p^+,5_{\bar{q}}^-,6_r^+, 7_{\bar{r}}^-)= \\
&+\frac{\la 57 \ra^2 \la 2 \ve 3+4 \ve 1]^2}
{s_{234} s_{567} \la 23 \ra \la 34 \ra \la 67 \ra \la 5 \ve 6+7 \ve 1]} \\
&+\frac{\la 25 \ra^2 [16]^2}{s_{671} 
\la 23 \ra \la 34 \ra [67] \la 5 \ve 6+7 \ve 1]} \, ,\\
\end{split}
\eeq
while $B_2$ is obtained by flipping
\beq
B_2=A_7^{(0)}(1_q^+,2_{\bar{p}}^-,3^+,4_p^+,5_{\bar{q}}^-,6_r^-, 7_{\bar{r}}^+)=
 -B_1({\rm flip}\,  6 \leftrightarrow 7 ) \, .
\eeq

Choosing $j=5, l=6$, the calculation for $B_3$ yields
\beq
\begin{split}
B_3&=A_7^{(0)}(1_q^+,2_{\bar{p}}^+,3^+,4_p^-,5_{\bar{q}}^-,6_r^+, 7_{\bar{r}}^-)= \\
&+\frac{\la 57 \ra^2 \la 4 \ve 3+2 \ve 1]^2}
{s_{234} s_{567} \la 23 \ra \la 34 \ra \la 67 \ra \la 5 \ve 6+7 \ve 1]} \\
&+\frac{\la 45 \ra^2 [16]^2}{s_{671} 
\la 23 \ra \la 34 \ra [67] \la 5 \ve 6+7 \ve 1]} \, ,\\
\end{split}
\eeq
and $B_4$ follows by performing the same flip as before
\beq
B_4=A_7^{(0)}(1_q^+,2_{\bar{p}}^+,3^+,4_p^-,5_{\bar{q}}^-,6_r^-, 7_{\bar{r}}^+)= 
 -B_3({\rm flip}\, , 6 \leftrightarrow 7 ) \, .
\eeq

$B_5$ has been computed by selecting $j=7, l=1$,
\beq
\begin{split}
B_5&=A_7^{(0)}(1_q^+,2_{\bar{p}}^-,3_p^+,4_{\bar{q}}^-,5^+,6_r^+, 7_{\bar{r}}^-)= \\
&+\frac{\la 46 \ra \la 47 \ra^2 [13]^2}{s_{123} [23] \la 45 \ra 
\la 56 \ra \la 67 \ra \la 4 \ve 3+2 \ve 1]} \\
&+\frac{\la 24 \ra^2 \la 6 \ve 5+7 \ve 1] \la 7 \ve 6+5 \ve 1]^2}{s_{234} s_{567} 
\la 23 \ra \la 56 \ra \la 67 \ra \la 4 \ve 3+2 \ve 1] \la 5 \ve 6+7 \ve 1]} \\
&+\frac{\la 24 \ra^2 [16]^2}
{s_{671} \la 23 \ra \la 45 \ra [67] \la 5 \ve 6+7 \ve 1]} \, ,\\
\end{split}
\eeq
and $B_6$ by taking $j=4, l=5$,
\beq
\begin{split}
B_6=&A_7^{(0)}(1_q^+,2_{\bar{p}}^-,3_p^+,4_{\bar{q}}^-,5^+,6_r^-, 7_{\bar{r}}^+)= \\
&-\frac{[71]^2 \la 24 \ra^2 \la 46 \ra^3}{\la 23 \ra \la 45 \ra \la 56 \ra
\la 4 \ve 3+2 \ve 1] \la 4 \ve 5+6 \ve 7] \la 6 \ve (7+1) (2+3) \ve 4 \ra} \\
&+\frac{[13]^2 \la 46 \ra^3}
{s_{123} [23] \la 45 \ra \la 56 \ra \la 67 \ra \la 4 \ve 3+2 \ve 1]} \\
&+\frac{\la 24 \ra^2 [57]^3}{s_{567} \la 23 \ra [67] 
\la 1 \ve 7+6 \ve 5] \la 4 \ve 5+6 \ve 7]} \\
&+\frac{\la 24 \ra^2 \la 6 \ve 7+1 \ve 5]^3}{s_{671} s_{234} \la 23 \ra 
\la 67 \ra \la 1 \ve 7+6 \ve 5] \la 6 \ve (7+1) (2+3) \ve 4 \ra } \, .\\
\end{split}
\eeq

$B_7$ can be obtained from $B_5$ by a flip operation,
\beq
B_7=A_7^{(0)}(1_q^+,2_{\bar{p}}^+,3_p^-,4_{\bar{q}}^-,5^+,6_r^+, 7_{\bar{r}}^-)=
 -B_5({\rm flip}\, , 2 \leftrightarrow 3 ) \, ,
\eeq
as can $B_8$ from $B_6$,
\beq
B_8=A_7^{(0)}(1_q^+,2_{\bar{p}}^+,3_p^-,4_{\bar{q}}^-,5^+,6_r^-, 7_{\bar{r}}^+)=
 -B_6({\rm flip}\, 2 \leftrightarrow 3 ) \, .
\eeq

We have computed $B_9$ by selecting $j=1, l=3$,
\beq
\begin{split}
B_9&=A_7^{(0)}(1_q^-,2_{\bar{p}}^-,3_p^+,4_{\bar{q}}^+,5^+,6_r^+, 7_{\bar{r}}^-)= \\
&+\frac{\la 12 \ra^2 \la 2 \ve 4+5 \ve 6]^2 }{\la 23 \ra \la 45 \ra [67]
\la 5 \ve (6+7) (1+3) \ve 2 \ra \la 2 \ve (4+5) (6+7) \ve 1 \ra} \\
&+\frac{\la 12 \ra^2 \la 7 \ve 6+5 \ve 4]^2 \la 6 \ve (5+7) (1+3) \ve 2 \ra}
{s_{567} s_{123} \la 23 \ra \la 56 \ra \la 67 \ra      
\la 1 \ve 2+3 \ve 4] \la 5 \ve (6+7) (1+3) \ve 2 \ra} \\
&+\frac{\la 17 \ra^2 [34]^2 \la 6 \ve 2+4 \ve 3]}{s_{234} [23] 
\la 56 \ra \la 67 \ra \la 5 \ve 4+2 \ve 3] \la 1 \ve 2+3 \ve 4]} \\
&+\frac{\la 17 \ra^2 \la 2 \ve 5+4 \ve 3]^2}{s_{671} \la 45 \ra \la 67 \ra
\la 5 \ve 4+2 \ve 3] \la 2 \ve (4+5) (6+7) \ve 1 \ra} \, .\\
\end{split}
\eeq

The calculation of $B_{10}$ has been performed by choosing $j=2, l=3$,
\beq
\begin{split}
B_{10}&=A_7^{(0)}(1_q^-,2_{\bar{p}}^-,3_p^+,4_{\bar{q}}^+,5^+,6_r^-, 7_{\bar{r}}^+)= \\
&+ \frac{\la 46 \ra \la 16 \ra^2 \la 6 \ve 5+4 \ve 3]^2}
{[23] \la 45 \ra \la 56 \ra \la 67 \ra 
\la 1 \ve (2+3) (4+5) \ve 6 \ra \la 6 \ve (7+1) (2+3) \ve 4 \ra} \\
&+ \frac{\la 16 \ra^2 \la 6 \ve 7+1 \ve 5] \la 2 \ve 3+4 \ve 5]^2} {s_{671} 
s_{234} \la 23 \ra \la 67 \ra \la 1 \ve 7+6 \ve 5] \la 6 \ve (7+1) (2+3) \ve 4 \ra} \\
&+ \frac{\la 12 \ra^2 [57]^3}
{s_{567} \la 23 \ra [67] \la 4 \ve 5+6 \ve 7] \la 1 \ve 7+6 \ve 5]} \\
&- \frac{\la 46 \ra \la 12 \ra^2 \la 6 \ve 5+4 \ve 7]^2}
{s_{123} \la 23 \ra \la 45 \ra \la 56 \ra \la 4 \ve 5+6 \ve 7]    
\la 1 \ve (2+3) (4+5) \ve 6 \ra} \, .\\
\end{split}
\eeq

Finally, a flip operation relates $B_{11}$ to $B_9$,
\beq
B_{11}=A_7^{(0)}(1_q^-,2_{\bar{p}}^+,3_p^-,4_{\bar{q}}^+,5^+,6_r^+, 7_{\bar{r}}^-)=
 -B_9({\rm flip}\, 2 \leftrightarrow 3 ) \, ,
 \eeq
and $B_{12}$ to $B_{10}$
\beq
B_{12}=A_7^{(0)}(1_q^-,2_{\bar{p}}^+,3_p^-,4_{\bar{q}}^+,5^+,6_r^-, 7_{\bar{r}}^+)= 
 -B_{10}({\rm flip}\, 2 \leftrightarrow 3 ) \, .
\eeq
\subsection{Sub-sub-leading amplitudes}
Due to the flavor configuration of the sub-sub-leading amplitudes, it is equivalent to insert 
the gluon in the middle of any quark-antiquark pair.  
The independent amplitudes we have computed are
\beq
\begin{split}
C_1=A_7^{(0)}(1_q^+,2_{\bar{p}}^-,3^+,4_p^+,5_{\bar{r}}^-,6_r^+, 7_{\bar{q}}^-) \qquad \qquad
&C_3=A_7^{(0)}(1_q^+,2_{\bar{p}}^+,3^+,4_p^-,5_{\bar{r}}^-,6_r^+, 7_{\bar{q}}^-) \\
C_2=A_7^{(0)}(1_q^+,2_{\bar{p}}^-,3^+,4_p^+,5_{\bar{r}}^+,6_r^-, 7_{\bar{q}}^-) \qquad \qquad
&C_4=A_7^{(0)}(1_q^-,2_{\bar{p}}^-,3^+,4_p^+,5_{\bar{r}}^-,6_r^+, 7_{\bar{q}}^+) \, .\\
\end{split}
\eeq

We were able to compute $C_1$, by picking $j=2, l=3$, as
\beq
\begin{split}
C_1&=A_7^{(0)}(1_q^+,2_{\bar{p}}^-,3^+,4_p^+,5_{\bar{r}}^-,6_r^+, 7_{\bar{q}}^-)= \\
&- \frac{\la 25 \ra^2 [61]^2 \la 2 \ve 6+7 \ve 1]}{s_{671} \la 23 \ra \la 34 \ra
[71] \la 2 \ve 1+7 \ve 6] \la 5 \ve 6+7 \ve 1]} \\
&- \frac{\la 57 \ra^2 \la 2 \ve 3+4 \ve 1]^2 \la 2 \ve (3+4) (6+7) \ve 5 \ra}
{s_{234} s_{567} \la 23 \ra \la 34 \ra \la 56 \ra \la 5 \ve 6+7 \ve 1]  
\la 2 \ve (3+4) (5+6) \ve 7 \ra} \\
&- \frac{\la 27 \ra^3 \la 4 \ve 3+2 \ve 6] \la 2 \ve 3+4 \ve 6]^2}{\la 71 \ra
\la 23 \ra \la 34 \ra [56] \la 2 \ve 1+7 \ve 6]  
\la 4 \ve (5+6) (7+1) \ve 2 \ra \la 2 \ve (3+4) (5+6) \ve 7 \ra } \\
&- \frac{\la 24 \ra \la 27 \ra^2 [16] \la 2 \ve 3+4 \ve 6]^2}
{\la 23 \ra \la 34 \ra [56] \la 2 \ve 1+7 \ve 6]  
\la 4 \ve (5+6) (7+1) \ve 2 \ra \la 2 \ve (3+4) (5+6) \ve 7 \ra } \\
&- \frac{\la 27 \ra^2 \la 5 \ve 6+4 \ve 3]^2}
{s_{456} s_{712} \la 56 \ra \la 71 \ra \la 4 \ve (5+6) (7+1) \ve 2 \ra} \, . \\
\end{split}
\eeq

Using the same tag, the amplitude $C_2$ is given by
\beq
\begin{split}
C_2&=A_7^{(0)}(1_q^+,2_{\bar{p}}^-,3^+,4_p^+,5_{\bar{r}}^+,6_r^-, 7_{\bar{q}}^-)= \\
&- \frac{\la 27 \ra^3 \la 4 \ve 3+2 \ve 6] \la 2 \ve 3+4 \ve 5]^2}
{\la 71 \ra \la 23 \ra \la 34 \ra [56] \la 2 \ve 1+7 \ve 6]  
\la 4 \ve (5+6) (7+1) \ve 2 \ra \la 2 \ve (3+4) (5+6) \ve 7 \ra} \\
&- \frac{\la 24 \ra \la 27 \ra^2 [16] \la 2 \ve 3+4 \ve 5]^2}
{\la 23 \ra \la 34 \ra [56] \la 2 \ve 1+7 \ve 6]  
\la 4 \ve (5+6) (7+1) \ve 2 \ra \la 2 \ve (3+4) (5+6) \ve 7 \ra} \\
&- \frac{\la 2 \ve 6+7 \ve 1]^3}
{s_{671} \la 23 \ra \la 34 \ra [71] \la 2 \ve 1+7 \ve 6] \la 5 \ve 6+7 \ve 1]} \\
&- \frac{\la 67 \ra^2 \la 2 \ve 3+4 \ve 1]^2 \la 2 \ve (3+4) (6+7) \ve 5 \ra}
{s_{234} s_{567} \la 23 \ra \la 34 \ra \la 56 \ra \la 5 \ve 6+7 \ve 1]  
\la 2 \ve (3+4) (5+6) \ve 7 \ra} \\
&- \frac{\la 27 \ra^2 \la 6 \ve 5+4 \ve 3]^2}
{s_{456} s_{712} \la 56 \ra \la 71 \ra \la 4 \ve (5+6) (7+1) \ve 2 \ra} \, .\\
\end{split}
\eeq

$C_3$ was computed by setting $j=1, l=2$, leading to
\beq
\begin{split}
C_3&=A_7^{(0)}(1_q^+,2_{\bar{p}}^+,3^+,4_p^-,5_{\bar{r}}^-,6_r^+, 7_{\bar{q}}^-)= \\
&- \frac{\la 57 \ra^2 \la 4 \ve 3+2 \ve 1]^2 \la 2 \ve (3+4) (6+7) \ve 5 \ra }
{s_{234} s_{567} \la 23 \ra \la 34 \ra \la 56 \ra \la 5 \ve 6+7 \ve 1]
\la 2 \ve (3+4) (5+6) \ve 7 \ra} \\
&- \frac{\la 57 \ra^3 \la 7 \ve 1+2 \ve 3]^2}{\la 56 \ra \la 71 \ra 
\la 7 \ve 6+5 \ve 4] \la 2 \ve (3+4) (5+6) \ve 7 \ra \la 7 \ve (1+2) (3+4) \ve 5 \ra} \\
&- \frac{\la 7 \ve 4+5 \ve 6]^3}{s_{456} \la 71 \ra \la 23 \ra [56] 
\la 3 \ve 4+5 \ve 6]
 \la 7 \ve 6+5 \ve 4 \ra} \\
&+ \frac{\la 45 \ra^2 \la 7 \ve 1+2 \ve 6]^3}{s_{712} \la 34 \ra \la 71 \ra
\la 2 \ve 1+7 \ve 6] \la 3 \ve 4+5 \ve 6] \la 7 \ve (1+2) (3+4) \ve 5 \ra} \\
&- \frac{\la 45 \ra^2 [16]^2 \la 2 \ve 6+7 \ve 1]}{s_{671} \la 23 \ra \la 34 \ra
[71] \la 5 \ve 6+7 \ve 1] \la 2 \ve 1+7 \ve 6]} \, .\\
\end{split}
\eeq

Finally, $C_4$ was calculated by picking $j=1, l=2$,
\beq
\begin{split}
C_4&=A_7^{(0)}(1_q^-,2_{\bar{p}}^-,3^+,4_p^+,5_{\bar{r}}^-,6_r^+, 7_{\bar{q}}^+)= \\
&- \frac{\la 25 \ra^2 [67]^2 \la 2 \ve 6+7 \ve 1]}{s_{671} \la 23 \ra \la 34 \ra
[71] \la 2 \ve 1+7 \ve 6] \la 5 \ve 6+7 \ve 1]} \\
&- \frac{\la 2 \ve (3+4) (6+7) \ve 5 \ra^3}
{s_{234} s_{567} \la 23 \ra \la 34 \ra \la 56 \ra \la 5 \ve 6+7 \ve 1]  
\la 2 \ve (3+4) (5+6) \ve 7 \ra} \\
&- \frac{\la 27 \ra \la 12 \ra^2 \la 4 \ve 3+2 \ve 6] \la 2 \ve 3+4 \ve 6]^2}
{\la 71 \ra \la 23 \ra \la 34 \ra [56] \la 2 \ve 1+7 \ve 6]  
\la 4 \ve (5+6) (7+1) \ve 2 \ra \la 2 \ve (3+4) (5+6) \ve 7 \ra } \\
&- \frac{\la 24 \ra \la 12 \ra^2 [16] \la 2 \ve 3+4 \ve 6]^2}
{\la 23 \ra \la 34 \ra [56] \la 2 \ve 1+7 \ve 6]  
\la 4 \ve (5+6) (7+1) \ve 2 \ra \la 2 \ve (3+4) (5+6) \ve 7 \ra } \\
&- \frac{\la 12 \ra^2 \la 5 \ve 6+4 \ve 3]^2}{s_{456} s_{712} 
\la 56 \ra \la 71 \ra \la 4 \ve (5+6) (7+1) \ve 2 \ra } \, .\\
\end{split}
\eeq

\section{Conclusions}
\label{sec:conclu}
In this paper we presented all  the four-quark plus three-gluon and six-quark plus one-gluon 
tree level NMHV helicity amplitudes. They were computed 
using the BCFW recursion relations. 
With these results  the full set of 
tree level helicity amplitudes, up to seven partons,  becomes available. 
Our formulae have been tested in all possible collinear and soft limits, which 
provides a very strict check for the correctness of the amplitudes. 
\footnote{Strictly speaking they are correct up to terms that must vanish 
in all possible soft and collinear limits, which are very unlikely to exist.}.

These amplitudes are a main ingredient for the calculation of multijets cross-sections
in hadronic colliders. As expected, we have obtained very compact expressions, allowing 
for a more convenient implementation 
in computer codes than those coming from automatic tree level computational methods. 

\noindent {\bf  Acknowledgements.} 
This work has been partially supported by ANPCYT, UBACyT and CONICET.

\appendix

\section{Appendix: six-parton NMHV results}
For the sake of completeness, we provide in this appendix the explicit results for the 
four-quark plus two-gluon and the six-quark NMHV amplitudes. 

\subsection{NMHV four-quark plus two-gluon amplitudes}
According to the color structure, there are again two different contributions, 
the leading "A" and the sub-leading "B" amplitudes. The truly independent ones are
\beq
\label{aamp}
\begin{split}
A_1= A_6^{(0)}(1_q^+,2_{\bar{q}}^-,3^+,4^-,5_p^+, 6_{\bar{p}}^-) \qquad \qquad
&A_4= A_6^{(0)}(1_q^+,2_{\bar{q}}^-,3^+,4_p^+, 5_{\bar{p}}^-,6^-) \\
A_2= A_6^{(0)}(1_q^+,2_{\bar{q}}^-,3^-,4^+,5_p^+, 6_{\bar{p}}^-) \qquad \qquad
&A_5= A_6^{(0)}(1_q^+,2_{\bar{q}}^-,3^+,4_p^-, 5_{\bar{p}}^+,6^-) \\
A_3= A_6^{(0)}(1_q^+,2_{\bar{q}}^-,3^+,4^-,5_p^-, 6_{\bar{p}}^+) \qquad \qquad
&A_6= A_6^{(0)}(1_q^+,2_{\bar{q}}^-,3^-,4_p^-, 5_{\bar{p}}^+,6^+) \\
\end{split}
\eeq
for leading, and
\beq
\begin{split}
B_1= A_6^{(0)}(1_q^+,2_{\bar{p}}^-,3^+,4^-,5_p^+, 6_{\bar{q}}^-) \qquad \qquad
&B_5= A_6^{(0)}(1_q^+,2_{\bar{p}}^-,3^+,4_p^+, 5_{\bar{q}}^-,6^-) \\
B_2= A_6^{(0)}(1_q^+,2_{\bar{p}}^-,3^-,4^+,5_p^+, 6_{\bar{q}}^-) \qquad \qquad
&B_6= A_6^{(0)}(1_q^+,2_{\bar{p}}^+,3^+,4_p^-, 5_{\bar{q}}^-,6^-) \\
B_3= A_6^{(0)}(1_q^+,2_{\bar{p}}^+,3^+,4^-,5_p^-, 6_{\bar{q}}^-) \qquad \qquad
&B_7= A_6^{(0)}(1_q^+,2_{\bar{p}}^+,3^-,4_p^-, 5_{\bar{q}}^-,6^+) \\
B_4= A_6^{(0)}(1_q^+,2_{\bar{p}}^+,3^-,4^+,5_p^-, 6_{\bar{q}}^-) \qquad \qquad
\end{split}
\eeq
for the sub-leading color contribution.
All the other amplitudes can be obtained by the use of discrete symmetries. 
\subsubsection{Leading amplitudes}

The most compact results we obtain for the six independent helicity amplitudes are
\beq
\begin{split}
A_1&= A_6^{(0)}(1_q^+,2_{\bar{q}}^-,3^+,4^-,5_p^+, 6_{\bar{p}}^-)= \\
&- \frac{\la 2 \ve 1+6 \ve 3] \la 26 \ra^2 [35]^3}
{s_{612} \la 12 \ra [34] [45] \la 2 \ve 1+6 \ve 5] \la 6 \ve 1+2 \ve 3]} \\
&- \frac{[13]^3 \la 46 \ra^3}
{s_{123} [12] \la 56 \ra \la 4 \ve 2+3 \ve 1] \la 6 \ve 1+2 \ve 3]} \\
&- \frac{\la 4 \ve 6+1 \ve 5] [15]^2 \la 24 \ra^3}
{s_{234} \la 23 \ra \la 34 \ra [56] \la 2 \ve 6+1 \ve 5] \la 4 \ve 5+6 \ve 1]} \, ,
\end{split}
\eeq
\\
\beq
\begin{split}
A_2&= A_6^{(0)}(1_q^+,2_{\bar{q}}^-,3^-,4^+,5_p^+, 6_{\bar{p}}^-)= \\
&- \frac{\la 2 \ve 1+6 \ve 4] \la 26 \ra^2 [45]^2}
{s_{612} \la 12 \ra [34] \la 2 \ve 1+6 \ve 5] \la 6 \ve 1+2 \ve 3]} \\
&- \frac{\la 5 \ve 2+3 \ve 1] \la 6 \ve 1+3\ve 2] \la 6 \ve 2+3 \ve 1]^2}
{s_{123} [12] [23] \la 45 \ra \la 56 \ra \la 4 \ve 2+3 \ve 1] 
\la 6 \ve 1+2 \ve 3]} \\
&- \frac{\la 3 \ve 6+1 \ve 5] [15]^2 \la 23 \ra^2}
{s_{234} \la 34 \ra [56] \la 2 \ve 6+1 \ve 5] \la 4 \ve 5+6 \ve 1]} \, ,
\end{split}
\eeq
\\
\beq
\begin{split}
A_3&= A_6^{(0)}(1_q^+,2_{\bar{q}}^-,3^+,4^-,5_p^-, 6_{\bar{p}}^+)= \\
&+ \frac{\la 4 \ve 2+3 \ve 5] [16]^2 \la 24 \ra^3} {s_{234} \la 23 \ra 
\la 34 \ra [56] \la 4 \ve 3+2 \ve 1] \la 2 \ve 3+4 \ve 5]} \\
&- \frac{[35] \la 2 \ve 5+4 \ve 3]^3}
{s_{345} \la 12 \ra [34] [45] \la 2 \ve 3+4 \ve 5] \la 6 \ve 1+2 \ve 3]} \\
&- \frac{\la 46 \ra \la 45 \ra^2 [13]^3}
{s_{123} [12] \la 56 \ra \la 4 \ve 3+2 \ve 1] \la 6 \ve 1+2 \ve 3]} \, ,
\end{split}
\eeq
\\
\beq
\begin{split}
A_4&= A_6^{(0)}(1_q^+,2_{\bar{q}}^-,3^+,4_p^+, 5_{\bar{p}}^-,6^-)= \\
&+\frac{[35] [34]^2 \la 26 \ra^3}
{s_{612} \la 12 \ra  [45] \la 2 \ve 6+1 \ve 5] \la 6 \ve 1+2 \ve 3]} \\
&+ \frac{\la 46 \ra \la 56 \ra^2 [13]^3}
{s_{123} [12]  \la 45 \ra \la 4 \ve 3+2 \ve 1] \la 6 \ve 5+4 \ve 3]} \\
&- \frac{[15] \la 24 \ra \la 2 \ve 5+6 \ve 1]^2}
{ \la 23 \ra \la 34 \ra [56] [61] \la 2 \ve 6+1 \ve 5] \la 4 \ve 5+6 \ve 1]} \, ,
\end{split}
\eeq
\\
\beq
\begin{split}
A_5&= A_6^{(0)}(1_q^+,2_{\bar{q}}^-,3^+,4_p^-, 5_{\bar{p}}^+,6^-)= \\
&-\frac{\la 24 \ra^3 [15]^3}{\la 23 \ra \la 34 \ra [56] [61] 
\la 4 \ve 3+2 \ve 1] \la 2 \ve 3+4 \ve 5]} \\
&+\frac{\la 26 \ra^3 [35]^3}{s_{345} \la 12 \ra [45] 
\la 2 \ve 3+4 \ve 5] \la 6 \ve 5+4 \ve 3]} \\
&-\frac{\la 46 \ra^3 [13]^3}{s_{123} [12] \la 45 \ra 
\la 6 \ve 1+2 \ve 3] \la 4 \ve 3+2 \ve 1]}  \, ,
\end{split}
\eeq
\\
\beq
\begin{split}
A_6&= A_6^{(0)}(1_q^+,2_{\bar{q}}^-,3^-,4_p^-, 5_{\bar{p}}^+,6^+)= \\
&+\frac{[26][16]^2 \la 35 \ra \la 34 \ra^2}{s_{345} [12] \la 45 \ra 
\la 3 \ve 4+5 \ve6] \la 5 \ve 4+3 \ve 2]}  \\ 
&+\frac{[46] [56]^2 \la 13 \ra \la 23 \ra^2}{s_{123} \la 12 \ra [45]  
\la 3 \ve 4+5 \ve 6] \la 1 \ve 2+3 \ve 4]} \\ 
&-\frac{(s_{234})^2 \la 15 \ra [24]}{[23] [34] \la 56 \ra \la 61 \ra 
\la 1 \ve 2+3 \ve 4] \la 5 \ve 4+3 \ve 2]} \, .
\end{split}
\eeq
The first four are computed selecting the legs as $j=6$ and $l=1$, while for the last 
two the chosen tags are $j=2$, $l=3$ and $j=3$, $l=4$, respectively. 

\subsubsection{Sub-leading amplitudes}
The expressions for the sub-leading amplitudes are again shorter than the leading ones, reading 
\beq
\begin{split}
B_1&= A_6^{(0)}(1_q^+,2_{\bar{p}}^-,3^+,4^-,5_p^+, 6_{\bar{q}}^-)= \\
&-\frac{[15]^2 \la 24 \ra^3}{s_{234}\la 23 \ra \la 34 \ra [61] 
\la 2 \ve 3+4 \ve 5]} \\ 
&- \frac{\la 26 \ra^2 [35]^3}{s_{345} \la 61 \ra [34] [45] 
\la 2 \ve 3+4 \ve 5]} \, ,
\end{split}
\eeq
\\
\beq
\begin{split}
B_2&= A_6^{(0)}(1_q^+,2_{\bar{p}}^-,3^-,4^+,5_p^+, 6_{\bar{q}}^-)= \\
&-\frac{\la 35 \ra \la 3 \ve 4+5 \ve 1]^2}{s_{345}\la 34 \ra \la 45 \ra [61]
\la 5 \ve 4+3 \ve 2]}\\
&- \frac{[24] \la 6 \ve 2+3 \ve 4]^2}{s_{234} [23] [34] \la 61 \ra 
\la 5 \ve 4+3 \ve 2]} \, ,
\end{split}
\eeq
\\
\beq
\begin{split}
B_3&= A_6^{(0)}(1_q^+,2_{\bar{p}}^+,3^+,4^-,5_p^-, 6_{\bar{q}}^-)= \\
&-\frac{\la 24 \ra \la 4 \ve 3+2 \ve 1]^2}{s_{234}\la 23 \ra \la 34 \ra [61]
\la 2 \ve 3+4 \ve 5]}\\
&- \frac{[35] \la 6 \ve 5+4 \ve 3]^2}{s_{345} [34] [45] \la 61 \ra 
\la 2 \ve 3+4 \ve 5]} \, ,
\end{split}
\eeq
\\
\beq
\begin{split}
B_4&=A_6^{(0)}(1_q^+,2_{\bar{p}}^+,3^-,4^+,5_p^-, 6_{\bar{q}}^-)= \\
&-\frac{[12]^2\la 35 \ra^3}{s_{345} \la 34 \ra \la 45 \ra [61] 
\la 5 \ve 4+3\ve 2] } \\
&-\frac{\la 56 \ra^2 [24]^3}{s_{234} [23] [34] \la 61 \ra 
\la 5 \ve 4+3 \ve 2]} \, , \\
\end{split}
\eeq
\\
\beq
\begin{split}
B_5&= A_6^{(0)}(1_q^+,2_{\bar{p}}^-,3^+,4_p^+, 5_{\bar{q}}^-,6^-)= \\
&- \frac{\la 2 \ve 3+4 \ve 1]^2}{s_{234}\la 23 \ra \la 34 \ra [56] [61]} \, ,
\end{split}
\eeq
\\
\beq
\begin{split}
B_6&= A_6^{(0)}(1_q^+,2_{\bar{p}}^+,3^+,4_p^-, 5_{\bar{q}}^-,6^-)= \\
&-\frac{\la 4 \ve 5+6 \ve 1]^2}{s_{234}\la 23 \ra \la 34 \ra [56] [61]} \, ,
\end{split}
\eeq
\\
\beq
\begin{split}
B_7&= A_6^{(0)}(1_q^+,2_{\bar{p}}^+,3^-,4_p^-, 5_{\bar{q}}^-,6^+)= \\
&-\frac{\la 5 \ve 6+1 \ve 2]^2}{s_{234} [23] [34] \la 56 \ra  \la 61 \ra} \, .
\end{split}
\eeq
The chosen legs $(j,l)$ to apply the BCFW recursion relation were (2,3), (3,4), (4,3), (3,4), (2,3), (6,1) and (5,6), respectively.

\subsection{NMHV six-quark amplitudes}
\label{ap6q}

Even though there are three color structures for the six-quark amplitudes, due to the lack of 
gluons the sub-sub-leading contributions can be directly obtained from the leading ones just by 
the cyclic property of the amplitudes.
 Therefore, there are only six independent helicity amplitudes for this process, given by the 
following choice 
\beq\label{sqamp}
\begin{split}
A_1=A_6^{(0)}(1_q^+,2_{\bar{q}}^-,3_p^+,4_{\bar{p}}^-,5_r^+, 6_{\bar{r}}^-) \qquad \qquad
&B_2=A_6^{(0)}(1_q^+,2_{\bar{p}}^-,3_p^+,4_{\bar{q}}^-,5_r^-, 6_{\bar{r}}^+) \\
A_2=A_6^{(0)}(1_q^+,2_{\bar{q}}^-,3_p^+,4_{\bar{p}}^-,5_r^-, 6_{\bar{r}}^+) \qquad \qquad
&B_3=A_6^{(0)}(1_q^+,2_{\bar{p}}^+,3_p^-,4_{\bar{q}}^-,5_r^+, 6_{\bar{r}}^-) \\
B_1=A_6^{(0)}(1_q^+,2_{\bar{p}}^-,3_p^+,4_{\bar{q}}^-,5_r^+, 6_{\bar{r}}^-) \qquad \qquad
&B_4=A_6^{(0)}(1_q^+,2_{\bar{p}}^+,3_p^-,4_{\bar{q}}^-,5_r^-, 6_{\bar{r}}^+) \, .
\end{split}
\eeq
The two independent leading amplitudes were obtained 
selecting the legs as $j=2$ and $l=3$, reading
\beq
\begin{split}
A_1&=A_6^{(0)}(1_q^+,2_{\bar{q}}^-,3_p^+,4_{\bar{p}}^-,5_r^+, 6_{\bar{r}}^-)=\\
&+\frac{ \la 24 \ra^2 [15]^2 \la 4 \ve 3+2 \ve 5]}{s_{234} \la 34 \ra [56] 
\la 4 \ve 3+2 \ve 1] \la 2 \ve 3+4 \ve 5]} \\
&+\frac{ \la 26 \ra^2 [35]^2 \la 2 \ve 5+4 \ve 3]}{s_{345} \la 12 \ra [34] 
\la 6 \ve 1+2 \ve 3] \la 2 \ve 3+4 \ve 5]} \\
&-\frac{ \la 46 \ra^2 [13]^2 \la 6 \ve 3+2 \ve 1]}{s_{123} [12] \la 56 \ra 
\la 4 \ve 3+2 \ve 1] \la 6 \ve 1+2 \ve 3]} \, , 
\end{split}
\eeq
and
\beq
\begin{split}
A_2&=A_6^{(0)}(1_q^+,2_{\bar{q}}^-,3_p^+,4_{\bar{p}}^-,5_r^-, 6_{\bar{r}}^+)=\\
&+\frac{ \la 24 \ra^2 [61]^2 \la 4 \ve 3+2 \ve 5]}{s_{234} \la 34 \ra [56] 
\la 4 \ve 3+2 \ve 1] \la 2 \ve 3+4 \ve 5]} \\
&-\frac{\la 2 \ve 5+4 \ve 3]^3}{s_{345} \la 12 \ra [34] 
\la 2 \ve 3+4 \ve 5] \la 6 \ve 5+4 \ve 3]} \\
&+\frac{ \la 45 \ra^2 [13]^2 \la 6 \ve 3+2 \ve 1]}{s_{123} [12] \la 56 \ra 
\la 4 \ve 3+2 \ve 1] \la 6 \ve 5+4 \ve 3]} \, . \\
\end{split}
\eeq

The sub-leading amplitudes are simply given by 
\beq
\begin{split}
B_1&=A_6^{(0)}(1_q^+,2_{\bar{p}}^-,3_p^+,4_{\bar{q}}^-,5_r^+, 6_{\bar{r}}^-)=\\
&+\frac{\la 46 \ra^2 [13]^2}{s_{123} [23] \la 56 \ra \la 4 \ve 3+2 \ve 1]}\\
&-\frac{\la 24 \ra^2 [51]^2}{s_{234} \la 23 \ra [56] \la 4 \ve 3+2 \ve 1]}\, ,
\end{split}
\eeq
\\
\beq
B_2=A_6^{(0)}(1_q^+,2_{\bar{p}}^-,3_p^+,4_{\bar{q}}^-,5_r^-, 6_{\bar{r}}^+)=
 -B_1({\rm flip}\, 5 \leftrightarrow 6 ) \, ,
\eeq
\\
\beq
\begin{split}
B_3&=A_6^{(0)}(1_q^+,2_{\bar{p}}^+,3_p^-,4_{\bar{q}}^-,5_r^+, 6_{\bar{r}}^-)=\\
&-\frac{\la 3 \ve 2+1 \ve 5]^2}
{s_{123} \la 23 \ra [56] \la 1 \ve 2+3 \ve 4]} \\
&+\frac{\la 6 \ve 4+3 \ve 2]^2}
{s_{234} [23] \la 56 \ra \la 1 \ve 2+3 \ve 4]} \, , \\
\end{split}
\eeq
and
\beq
B_4=A_6^{(0)}(1_q^+,2_{\bar{p}}^+,3_p^-,4_{\bar{q}}^-,5_r^-, 6_{\bar{r}}^+)=
 -B_3({\rm flip}\, 5 \leftrightarrow 6 ) \, .
\eeq

\end{document}